\documentclass[twocolumn,showpacs,english,superscriptaddress,preprintnumbers,amsmath,amssymb,floatfix]{revtex4-1}
\usepackage[T1]{fontenc}
\usepackage[latin9]{inputenc}
\usepackage{dcolumn}
\usepackage{bm}
\usepackage{graphicx}
\usepackage{color}
\usepackage{esint}
\usepackage{babel}
\usepackage{amsfonts}
\usepackage{slashed}
\usepackage{xspace}
\usepackage{amsmath}

\def\be{\begin{equation}}
\def\ee{\end{equation}}
\def\bea{\begin{eqnarray}}
\def\eea{\end{eqnarray}}
\def\bmat{\begin{pmatrix}}
\def\emat{\end{pmatrix}}
\def\bs{\begin{split}}
\def\es{\end{split}}

\def\~{$\approx$}
\def\bra{\langle}
\def\ket{\rangle}
\def\up{\uparrow}
\def\dw{\downarrow}
\def\dag{\dagger}

\begin{document}

\title{Josephson effect in junctions of conventional and topological superconductors}

\author{A.~Zazunov}
\affiliation{Institut f\"ur Theoretische Physik, Heinrich-Heine-Universit\"at, D-40225  D\"usseldorf, Germany}

%\author[1]{A.~Iks}
\author{A.~Iks}
\affiliation{Institut f\"ur Theoretische Physik, Heinrich-Heine-Universit\"at, D-40225  D\"usseldorf, Germany}

\author{M.~Alvarado}
\affiliation{Departamento de F{\'i}sica Te{\'o}rica de la Materia Condensada C-V,
Condensed Matter Physics Center (IFIMAC) and Instituto Nicol\'as  Cabrera,
 Universidad Aut{\'o}noma de Madrid, E-28049 Madrid, Spain}

%\author[2]{A.~Levy Yeyati}
\author{A.~Levy Yeyati}
\affiliation{Departamento de F{\'i}sica Te{\'o}rica de la Materia Condensada C-V, Condensed Matter Physics Center (IFIMAC) and Instituto Nicol\'as  Cabrera,  Universidad Aut{\'o}noma de Madrid, E-28049 Madrid, Spain}
%\author*[1]{R.~Egger}{egger@hhu.de}
\author{R.~Egger}
\affiliation{Institut f\"ur Theoretische Physik, Heinrich-Heine-Universit\"at, D-40225  D\"usseldorf, Germany}

\date{\today}
%\maketitle

\begin{abstract}
 We present a theoretical analysis of the equilibrium Josephson current-phase relation 
 in hybrid devices made of conventional $s$-wave spin-singlet superconductors (S) and 
 topological superconductor (TS) wires featuring Majorana end states.  Using Green's function
 techniques, the topological superconductor is alternatively described by the low-energy continuum limit of a Kitaev chain or by a more microscopic spinful nanowire model.  
 We show that for the simplest S-TS tunnel junction, only the $s$-wave pairing correlations
 in a spinful TS nanowire model can generate a Josephson effect.
 The critical current is much smaller in the topological regime and exhibits a kink-like  dependence on the Zeeman field along the wire.
  When a correlated quantum dot (QD) in the magnetic regime is 
  present in the junction region, however, the Josephson current becomes finite also in the deep topological phase
   as shown for the cotunneling regime and by a mean-field analysis.
 Remarkably, we find that the S-QD-TS setup can support $\varphi_0$-junction behavior, where a 
 finite supercurrent flows at vanishing phase difference. 
 Finally, we also address a multi-terminal S-TS-S geometry, where the TS wire acts as 
 tunable parity switch on the Andreev bound states in a superconducting atomic contact.  
\end{abstract}
\maketitle

\section{Introduction}

The physics of topological superconductors (TSs) is being vigorously explored at present. After Kitaev \cite{Kitaev2001} showed that a one-dimensional  (1D)
spinless fermionic lattice model with nearest-neighbor $p$-wave pairing (`Kitaev chain') features a topologically nontrivial phase with Majorana bound states (MBSs) at open boundaries, 
Refs.~\cite{Lutchyn2010,Oreg2010} have pointed out that the physics of the Kitaev chain
could be realized in spin-orbit coupled nanowires with a magnetic Zeeman field and in
the proximity to a nearby $s$-wave superconductor.  The spinful nanowire model of Refs.~\cite{Lutchyn2010,Oreg2010} indeed features $p$-wave pairing correlations for appropriately 
chosen model parameters. In addition, it also 
contains $s$-wave pairing correlations which become gradually smaller as one moves into 
the deep topological regime. Topologically nontrivial hybrid semiconductor nanowire devices 
are of considerable interest in the context of quantum information processing  \cite{Alicea2011,Alicea2012,Leijnse2012,Beenakker2013,Aasen2016,Landau2016,Plugge2016,Plugge2017,Aguado2017}, and they may also be designed in 
 two-dimensional layouts by means of gate lithography techniques. 
 Over the last few years, several experiments employing such platforms have provided 
mounting evidence for MBSs, e.g., from zero-bias conductance peaks in N-TS junctions 
(where N stands for a normal-conducting lead) and via signatures of the $4\pi$-periodic Josephson effect in TS-TS junctions
 \cite{Mourik2012,Das2012,Albrecht2016,Deng2016,Guel2017,Albrecht2017,Zhang2017,
 Nichele2017,Suominen2017,Gazi2017,Zhang2018,Deng2018,Laroche2018}.   
 Related MBS phenomena have  been reported for other material platforms as well,
 see, e.g., Refs.~\cite{Yazdani2014,Franke2015,Sun2016,Feldman2017,Deacon2017}, and 
 most of the results reported below also apply to those settings.
 Available materials are often of sufficiently high quality to meet the conditions for ballistic transport, 
 and we will therefore neglect disorder effects.

 In view of the large amount of published theoretical works on the Josephson effect in such 
 systems, let us first motivate the present study.  (For a more detailed discussion and references, 
 see below.)  Our manuscript addresses the supercurrent flowing in Josephson junctions with a 
magnetic impurity. By considering Josephson junctions between a topological superconductor 
and a non-topological superconductor, we naturally extend previous works on 
Josephson junctions with a magnetic impurity between two conventional superconductors, as well as other works on Josephson junctions between topological and non-topological superconductors but without a magnetic impurity. In the simplest
description, Josephson junctions between topological and non-topological supeconductors carry no
supercurrent. Instead, a supercurrent can flow only with certain deviations from the idealized model
description. The presence of a magnetic impurity in the junction is one of these deviations, and 
this effect allows for novel signatures for the topological transition via the so-called $\varphi_0$ behavior and/or through the kink-like dependence of the critical current on a Zeeman field driving the transition.
We consider two different geometries in various regimes, e.g., the cotunneling regime where a controlled perturbation theory is possible, and a 
mean-field description of the stronger-coupling regime. We study both
idealized Hamiltonians (allowing for analytical progress) as well as more realistic models 
for the superconductors.

\begin{figure}
\hspace*{-0.6cm}
\includegraphics[width=1.3\columnwidth]{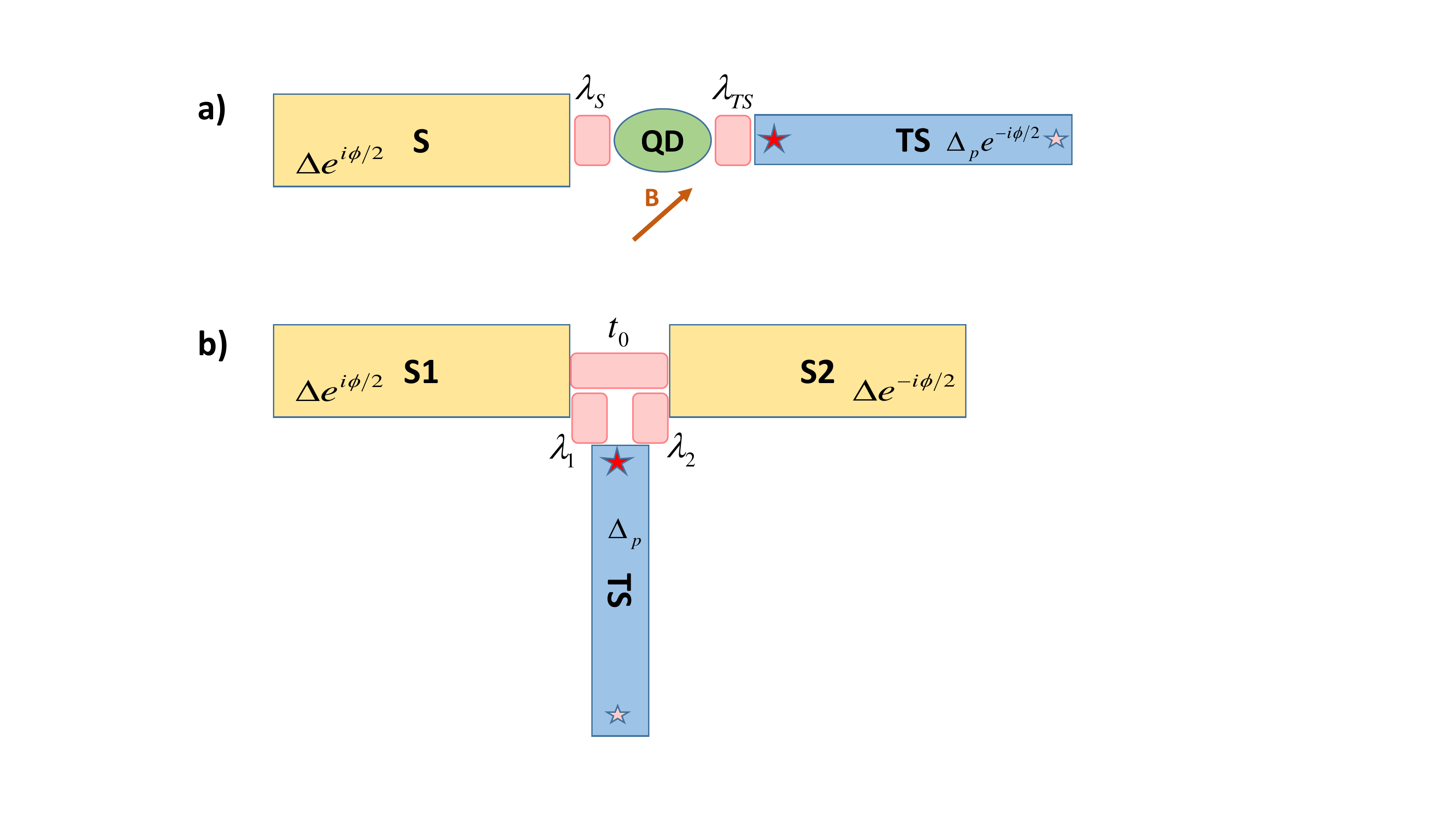}
\vspace*{-1cm}
\caption{Schematic setups studied in this paper. {\bf a)} S-QD-TS geometry:
S denotes a conventional $s$-wave BCS superconductor with order parameter 
$\Delta e^{i\phi/2}$, and TS represents a topologically nontrivial superconducting wire with MBSs (shown as stars) and proximity-induced order parameter 
$\Delta_p e^{-i\phi/2}$.  The interface contains
a quantum dot (QD) corresponding to an Anderson impurity,  connected to the S/TS leads by tunnel amplitudes $\lambda_{S/TS}$ (light red).  
The QD is also exposed to a local Zeeman field ${\bf B}$.
{\bf b)} S-TS-S geometry: Two conventional superconductors (S1 and S2) with the same gap $\Delta$ and a TS wire with proximity gap $\Delta_p$  form a  trijunction. The  order parameter phase
 of S1 (S2), $\phi_1=\phi/2$ ($\phi_2=-\phi/2$), is taken relative to the phase of the TS wire, and tunnel couplings $\lambda_{1/2}$ connect S1/S2 to the TS wire.
When the TS wire is decoupled ($\lambda_{1,2}=0$), the S-S junction becomes a standard SAC with transparency ${\cal T}$ determined by the tunnel amplitude $t_0$, see Eq.~(\ref{tau0}). 
}
\label{fig1}
\end{figure}

 To be more specific, we address the equilibrium current-phase relation (CPR)  
 in different setups involving both conventional $s$-wave BCS superconductors (`S' leads) 
 and TS wires, see Fig.~\ref{fig1} for a schematic 
 illustration. In general, the CPR is closely related to the Andreev bound state (ABS) spectrum of
  the system.   For S-TS junctions with the TS wire deep in the topological phase such that it can be modeled by a Kitaev chain, the supercurrent vanishes
  identically \cite{Zazunov2012}.  This supercurrent blockade can be traced back to the
  different ($s/p$-wave) pairing symmetries for the S/TS leads, together with the fact that 
  MBSs have a definite spin polarization. For an early study of Josephson currents between superconductors with different ($p/d$) pairing symmetries, see also Ref.~\cite{Kwon2004}.
   A related phenomenon concerns Multiple Andreev Reflection (MAR) features in nonequilibrium superconducting quantum transport at subgap voltages
    \cite{Bratus1995,Averin1995,Cuevas1996,Nazarov2009}.  Indeed, it has been established that MAR processes are absent in S-TS junctions (with the TS wire in the deep topological regime) such that
    only quasiparticle transport above the gap is possible \cite{Peng2015,Ioselevich2016,Sharma2016,Zazunov2016,Setiawan2017a,Setiawan2017b,Zazunov2017,Sen2017}.
   
   There are several ways to circumvent this supercurrent blockade in S-TS junctions.
(i) One possibility has been described in Ref.~\cite{Zazunov2017}. For a trijunction formed by two TS wires and one S lead, crossed Andreev reflections allow for the nonlocal
 splitting of Cooper pairs in the S electrode involving both TS wires (or the reverse process).  In this way,   
an equilibrium supercurrent will be generated unless the MBS spin polarization axes of both TS wires are precisely aligned.
(ii) Even for a simple S-TS junction, a finite Josephson current is expected
  when the TS wire is modeled as spinful nanowire. This effect is due to the residual $s$-wave pairing character of the spinful TS model \cite{Lutchyn2010,Oreg2010}. 
  Interestingly, upon changing a control parameter, e.g., the bulk Zeeman field, which drives the TS wire across the topological phase transition, 
  we find that the critical current exhibits a kink-like feature that is mainly caused by a suppression of the Andreev state contribution in the topological phase.
(iii) Yet another possibility is offered by junctions containing a magnetic impurity in a local magnetic field.  
We here analyze the S-QD-TS setup in Fig.~\ref{fig1}(a) in some detail, where a quantum
dot (QD) is present within the S-TS junction region.  The QD is
modeled as an Anderson impurity \cite{Nazarov2009}, which is equivalent to a spin-$1/2$ quantum impurity over
a wide parameter regime.  Once spin mixing is induced by the magnetic impurity and the local magnetic field,
we predict that a finite Josephson current flows even in the deep topological limit. In particular,
in the cotunneling regime, we find an anomalous Josephson effect with finite supercurrent at vanishing phase difference 
($\varphi_0$-junction behavior) 
\cite{Buzdin2008,Zazunov2009,Schrade2017},   see also Refs.~\cite{Yokoyama2014,Camjayi2017,Cayao2017,Cayao2018}. 
 The $2\pi$-periodic CPR found
in S-QD-TS junctions could thereby provide independent evidence for MBSs via the anomalous
Josephson effect.  In addition, we compute the CPR within the mean-field approximation in order to go
beyond perturbation theory in the tunnel couplings connecting the QD to the superconducting leads.  Our mean-field
analysis shows that the $\varphi_0$-junction behavior is a generic feature for S-QD-TS devices in the topological regime which is not limited to the cotunneling regime.

 In the final part of the paper, we turn to the three-terminal 
 S-TS-S setup shown in Fig.~\ref{fig1}(b), 
 where the S-S junction by itself (with the TS wire decoupled) represents a standard 
 superconducting atomic contact  (SAC) with variable transparency of the weak link. 
 Recent experiments have demonstrated that the many-body ABS configurations of a 
 SAC can be probed and manipulated to high accuracy by microwave 
 spectroscopy \cite{Brethreau2013a,Brethreau2013b,Janvier2015}.  When the TS wire is coupled to
 the S-S junction, see Fig.~\ref{fig1}(b), the Majorana end state acts 
 as a parity switch on the ABS system of the SAC.  This effect allows for additional
 functionalities in Andreev spectroscopy.
We note that similar ideas have also been explored for TS-N-TS systems \cite{Tarasinski2015}.

\section{Results and Discussion}

\subsection{S-QD-TS junction}

\subsubsection{Model}

Let us start with the case of an S-QD-TS junction, where an interacting spin-degenerate single-level quantum dot (QD) is sandwiched between a conventional $s$-wave superconductor (S) and a topological superconductor (TS).  This geometry is shown in Fig.~\ref{fig1}(a).
The corresponding topologically trivial S-QD-S problem has been studied in great detail over the
past decades both theoretically
\cite{Glazman1989,Rozhkov1999,Vecino2003,Siano2004,Choi2004,Karrasch2008,Alvaro2011,Luitz2012}
and experimentally \cite{Kasumov1999,VanDam2006,Cleuziou2006,Jorgensen2007,Eichler2009,Delagrange2015}.  A main motivation for those studies came from the fact that the QD can be driven into the magnetic regime where it represents a spin-$1/2$ impurity subject to Kondo screening by the leads.  
The Kondo effect then competes against the superconducting 
bulk gap and one encounters local quantum phase transitions. 
By now,  good agreement between experiment and theory has been established.   
Rather than studying the fate of the Kondo effect in the S-QD-TS setting of 
Fig.~\ref{fig1}(a), we here pursue two more modest goals. First, 
we shall discuss the cotunneling regime in detail, where one can employ
perturbation theory in the dot-lead couplings. This regime exhibits
$\pi$-junction behavior in the S-QD-S case \cite{Glazman1989}. 
Second, in order to go beyond the cotunneling regime, we have 
performed a mean-field analysis similar in spirit to earlier work for S-QD-S devices \cite{Rozhkov1999,Vecino2003}. 

The Hamiltonian for the setup in Fig.~\ref{fig1}(a) is given by
\begin{equation}\label{htot}
H = H_{\rm S} + H_{\rm TS} + H_{\rm QD} + H_{\rm tun},
\end{equation}
where $H_{\rm S/TS}$ and $H_{\rm QD}$ describe the semi-infinite S/TS leads and the isolated dot in between, respectively,
 and $H_{\rm tun}$ refers to the tunnel contacts. We often use units with $e= \hbar = k_B=1$,  and $\beta=1/T$ denotes inverse temperature.
The QD is modeled as an Anderson impurity \cite{Nazarov2009}, i.e., a 
single spin-degenerate level of energy $\epsilon_0$ with repulsive on-site interaction 
energy $U>0$,
\begin{equation}\label{QDeq}
H_{\rm QD} = \sum_{\sigma = \uparrow, \downarrow} \epsilon_0 \left( n_\sigma -\frac12 \right) + U n_\uparrow n_\downarrow - {\bf B \cdot S} ,
\end{equation}
where the QD occupation numbers are $n_\sigma = d^\dagger_\sigma d_\sigma=0,1$,
with dot fermion operators $d_\sigma$ and $d^\dagger_\sigma$ for spin $\sigma$.  Using
standard Pauli matrices $\sigma_{x,y,z}$, we define 
\begin{equation}\label{sdef}
{\bf S}_{i = x,y,z} =\sum_{\sigma, \sigma'} d^\dagger_\sigma 
\left( \sigma_i \right)_{\sigma \sigma'} d_{\sigma'},
\end{equation}
such that ${\bf S}/2$ is a spin-1/2 operator.
In the setup of Fig.~\ref{fig1}(a), we also take into account an 
external Zeeman field ${\bf B} = (B_x, B_y, B_z)$ acting on the QD spin, 
where the units in Eq.~(\ref{QDeq}) include gyromagnetic and Bohr magneton factors.
The spinful nanowire proposal for TS wires \cite{Lutchyn2010,Oreg2010}
also requires a sufficiently strong bulk Zeeman field oriented along the wire in order to realize the topologically nontrivial phase, 
but for concreteness, we here imagine the field ${\bf B}$ as independent local field coupled
only to the QD spin.   One could use, e.g., a ferromagnetic grain near the QD to generate it. This field here
plays a crucial role because for ${\bf B}=0$, the S$+$QD part is spin rotation [SU(2)]
invariant and the arguments of Ref.~\cite{Zazunov2012} then rule out a supercurrent for TS wires in the deep topological regime.
We show below that unless ${\bf B}$ is inadvertently aligned with the
MBS spin polarization axis,  spin mixing will indeed generate a supercurrent.

The S/TS leads are coupled to the QD via a tunneling Hamiltonian \cite{Zazunov2011},
\begin{equation} \label{ht}
H_{\rm tun}  = \lambda_S \sum_{\sigma = \uparrow, \downarrow} \psi_\sigma^\dagger d_\sigma + \lambda_{TS} e^{-i \phi/2} \psi^\dagger d_\uparrow + {\rm h.c.},
\end{equation}
where $\psi_\sigma$ and $\psi$ are boundary fermion fields representing the S lead and the effectively spinless TS lead, respectively. 
For the S lead, we assume the usual BCS model \cite{Alvaro2011}, where the operator $\psi_\sigma$ annihilates 
an electron with spin $\sigma$ at the junction.
 The TS wire will, for the moment, be described by the low-energy Hamiltonian of a Kitaev chain 
 in the deep topological phase with chemical potential $\mu=0$ \cite{Kitaev2001,Alicea2012}.  
 The corresponding fermion operator $\psi$ at the junction includes both the MBS contribution and above-gap quasiparticles \cite{Zazunov2016}.
Without loss of generality, we choose the unit vector $\hat e_z$ as the MBS spin polarization direction 
and take real-valued tunnel amplitudes $\lambda_{S/TS}$, see Fig.~\ref{fig1}(a), using a gauge where the 
superconducting phase difference $\phi$ appears via the QD-TS tunneling term.  These tunnel amplitudes contain density-of-states factors
for the respective leads.
The operator expression for the current flowing through the system is then given by
\begin{equation}\label{hatI}
\hat I = \frac{2e}{ \hbar} \partial_\phi H_{\rm tun} .
\end{equation}

We do not specify $H_{\rm S/TS}$ in Eq.~(\ref{htot}) explicitly since 
within the imaginary-time ($\tau$) boundary Green's function (bGF) formalism \cite{Zazunov2016} employed here,  
we only need to know the bGFs.  For the S lead with
 gap value $\Delta$, the bGF has the Nambu matrix form \cite{Zazunov2016}
\begin{eqnarray}\label{g}
g(\tau) &=& - \langle {\cal T}_\tau \Psi_{\rm S}(\tau) \Psi^\dagger_{\rm S}(0) \rangle_0^{} = \beta^{-1} \sum_\omega e^{-i \omega \tau} g(\omega) ,\\ \nonumber
\Psi_{\rm S} &=& \left( \begin{array}{c} \psi_\uparrow \\  \psi^\dagger_\downarrow \end{array} \right) ,\quad g(\omega) = - \frac{i \omega \tau_0 + \Delta \tau_x}{\sqrt{\omega^2 + \Delta^2}} ,
\end{eqnarray}
where the expectation value $\langle \cdots \rangle_0$ refers to an isolated S lead,
 ${\cal T}_\tau$ denotes time ordering, $\omega$ runs over fermionic Matsubara frequencies,
 i.e., $\omega=2\pi(n+1/2)/\beta$ with integer $n$, and we define Pauli (unity) matrices 
 $\tau_{x,y,z}$  ($\tau_0$)  in particle-hole space
 corresponding to the Nambu spinor $\Psi_{\rm S}$. 
Similarly, for a TS lead with proximity-induced gap $\Delta_p$, the low-energy limit of a 
Kitaev chain yields the bGF \cite{Zazunov2016} 
\begin{eqnarray} \nonumber
G(\tau) &=&- \langle {\cal T}_\tau \Psi_{\rm TS} (\tau) \Psi^\dagger_{\rm TS}(0) \rangle_0 ,\quad
\Psi_{\rm TS} = \left( \begin{array}{c} \psi \\ \psi^\dagger \end{array} \right) ,\\
G(\omega) &=& \frac {1}{ i \omega}\left( \sqrt{\omega^2 + \Delta_p^2} \, 
\tau_0 + \Delta_p \tau_x \right)  .\label{G}
\end{eqnarray}
The matrices $\tau_{0,x}$ here act in the Nambu space defined by the spinor $\Psi_{\rm TS}$.
 Later on we will address how our results change when the TS wire is modeled as spinful nanowire \cite{Lutchyn2010,Oreg2010}, where the corresponding bGF has been specified in 
 Ref.~\cite{Zazunov2017}.  We emphasize that the bGF (\ref{G})  captures the effects of both the MBS (via the $1/\omega$ term) and of the above-gap
  continuum quasiparticles (via the square root) \cite{Zazunov2016,Peng2017}. 
 
In most of the following discussion, we will assume that $U$ is the dominant energy scale, with the single-particle level located at
 $\epsilon_0 \approx - U /2$.  In that case, low-energy states with energy well below $U$ are restricted to the single occupancy sector,
\begin{equation}\label{constr}
n_\uparrow + n_\downarrow = 1 ,
\end{equation}
and the QD degrees of freedom become equivalent to the spin-1/2 operator ${\bf S}/2$ in Eq.~(\ref{sdef}). 
In this regime, the QD acts like a magnetic impurity embedded in the S-TS junction.
Using a Schrieffer-Wolff transformation to project the full Hamiltonian to the Hilbert subspace satisfying Eq.~(\ref{constr}), $H\to H_{\rm eff}$,
one arrives at the effective low-energy Hamiltonian
\begin{equation}\label{heff1}
H_{\rm eff} = H_0 + H_{\rm int} ,   \quad H_0 = H_{\rm S} + H_{\rm TS} - {\bf B \cdot S} ,
\end{equation}
with the interaction term
\begin{eqnarray} \nonumber
H_{\rm int} & = & -   \frac{2 }{ U} \sum_{\sigma , \sigma'}
\left( \eta_\sigma^\dagger d_\sigma d_{\sigma'}^\dagger \eta_{\sigma'}  + {\rm h.c.} \right)
\\ & = & \nonumber
\frac{2}{ U} \sum_{\sigma = \uparrow/\downarrow = \pm} \left(
\sigma S_z \eta^\dagger_\sigma \eta_\sigma + S_\sigma \eta^\dagger_{-\sigma} \eta_\sigma
\right)  \\ \label{Hint}
&+& \frac{2} {U} \delta n \sum_\sigma \eta^\dagger_\sigma \eta_\sigma - 
\frac{2 \Lambda}{ U} (\delta n+1),   
\end{eqnarray}
where $S_\pm = S_x \pm i S_y$  and $\delta n = \sum_\sigma n_\sigma - 1$. Moreover,
$\Lambda = \left[ \eta_\sigma , \eta^\dagger_\sigma \right]_+$ is the 
anticommutator of the composite boundary fields
\begin{equation}\label{eta}
\eta_\sigma = \lambda_S \psi_\sigma +  \delta_{\sigma, \uparrow} 
\lambda_{TS} e^{i \phi/2} \psi .
\end{equation}
We note that $\Lambda$ is real-valued and does not depend on $\phi$.
Due to the constraint (\ref{constr}) on the dot occupation, the last two terms in 
Eq.~(\ref{Hint}) do not contribute to the system dynamics and we obtain
\begin{eqnarray}\label{Hint2}
H_{\rm int} &=& \frac{4}{U}
 \sum_{\sigma, \sigma'} {\cal Q}_{\sigma \sigma'} \eta^\dagger_\sigma \eta_{\sigma'} ,\\ \nonumber
{\cal Q}_{\sigma \sigma} &=& \frac{\sigma}{2} S_z,  \quad  {\cal Q}_{\sigma, -\sigma} = 
\frac12 S_{-\sigma}.
\end{eqnarray}
A formally exact expression for the partition function is then given by
\begin{equation}\label{Z}
Z = {\rm Tr} \Big|_{\delta n = 0} \left(
e^{-\beta H_0} {\cal T}_\tau e^{-\int_0^\beta d\tau H_{\rm int}(\tau)} \right) ,
\end{equation}
where $H_{\rm int}(\tau) = e^{\tau H_0} H_{\rm int} e^{-\tau H_0}$ with $H_0$ in Eq.~(\ref{heff1}) and the trace 
extends only over the Hilbert subspace corresponding to Eq.~(\ref{constr}).
We can equivalently write Eq.~(\ref{Z}) in the form
\begin{eqnarray}\label{zz}
Z &=& Z_0\left \langle {\cal T}_\tau e^{- \beta \hat W} \right\rangle_0 = e^{- \beta F},\\
\hat W &=& \beta^{-1} \int_0^\beta d\tau H_{\rm int}(\tau) , \nonumber \\ \nonumber
Z_0 &=& {\rm Tr} \Big|_{\delta n = 0}  e^{-\beta H_0}  = e^{- \beta F_0},
\end{eqnarray}
where $F$ is the free energy.
The Josephson current then follows as $I =(2e/\hbar) \partial_\phi F$, see Eq.~(\ref{hatI}).

\subsubsection{Cotunneling regime}

We now address the CPR in the elastic cotunneling regime,
\begin{equation} 
\lambda_S\lambda_{TS}\ll {\rm min} \{\Delta,\Delta_p,U\},
\end{equation}
 where perturbation theory in $H_{\rm int}$ is justified. We thus wish to
compute the free energy $F(\phi)$ from Eq.~(\ref{zz}) to 
lowest nontrivial order.   
With $W_0 = \langle \hat W \rangle_0$, the standard cumulant expansion gives
\begin{equation}\label{F}
F - F_0 = W_0 - \frac{\beta}{2} \left( \left\langle \hat W^2 \right\rangle_0 - W_0^2 \right) +  {\cal O}(W^3).
\end{equation}
By virtue of Wick's theorem, time-ordered correlation functions of the boundary operators (\ref{eta})
are now expressed in terms of S/TS bGF matrix elements, see Eqs.~(\ref{g}) and (\ref{G}),
\begin{eqnarray}\nonumber
 \langle {\cal T}_\tau \eta_\sigma (\tau) \eta^\dagger_{\sigma'}(0) \rangle^{}_0 &=&
\delta_{\sigma \sigma'} \Big[
\lambda_S^2 \langle {\cal T}_\tau\psi_\sigma(\tau) \psi^\dagger_\sigma(0) \rangle^{}_0 + \\
&+&\label{etabareta}
\delta_{\sigma, \uparrow} \lambda_{TS}^2 \langle {\cal T}_\tau \psi(\tau) \psi^\dagger(0) 
\rangle^{}_0 \Big] 
\end{eqnarray}
and similarly
\begin{eqnarray}\label{etaeta}
\langle {\cal T}_\tau \eta_\sigma (\tau) \eta_{\sigma'}(0) \rangle_0^{} &=&
\delta_{\sigma, -\sigma'}
\lambda_S^2 \langle {\cal T}_\tau \psi_\sigma(\tau) \psi_{-\sigma}(0) \rangle^{}_0 \\
&+& e^{i \phi}
\delta_{\sigma \sigma'} \delta_{\sigma, \uparrow} \lambda_{TS}^2 
\langle {\cal T}_\tau \psi(\tau) \psi(0) \rangle^{}_0 .\nonumber
\end{eqnarray}
Next we observe that $\partial_\phi \langle H_{\rm int} \rangle_0 = 0$. As a consequence,
 the $\phi$-independent terms $W_0$ and $W_0^2$ in Eq.~(\ref{F}) do not contribute to the Josephson current. The leading contribution is then of second order in $H_{\rm int}$,
\begin{eqnarray}\nonumber
I(\phi) &=& - \beta^{-1} \partial_\phi \int_0^\beta d \tau_1 d \tau_2 \langle 
{\cal T}_\tau H_{\rm int}(\tau_1) H_{\rm int}(\tau_2) \rangle^{}_0 \\ \nonumber
& =& -\frac{\kappa^2}{\beta} \int_0^\beta d \tau_1 d \tau_2 \ g_{12}(\tau_1 - \tau_2) G_{12}(\tau_1 - \tau_2) \\
&\times& ie^{i \phi} \sum_\sigma \sigma \langle {\cal T}_\tau {\cal Q}_{\sigma, \uparrow}(\tau_1) {\cal Q}_{-\sigma, \uparrow}(\tau_2) \rangle^{}_0 +{\rm h.c.},\label{I2}
\end{eqnarray}
with ${\cal Q}_{\sigma,\sigma'}$ in Eq.~(\ref{Hint2}) and the small dimensionless parameter 
\begin{equation}\label{kappa}
\kappa=\frac{4\lambda_S\lambda_{TS}}{U} \ll 1.
\end{equation}
From Eqs.~(\ref{g}) and (\ref{G}), the bGF matrix elements needed in Eq.~(\ref{I2}) follow as
\begin{eqnarray}\label{g12}
g_{12}(\tau) &=& - \frac{\Delta}{ \beta} \sum_\omega \frac{\cos (\omega \tau)}{ \sqrt{\omega^2 + \Delta^2}}  ,\\
 \nonumber G_{12}(\tau) &=& - \frac{\Delta_p}{ \beta} \sum_\omega \frac{\sin (\omega \tau)}{ \omega} \simeq -  \frac{\Delta_p }{2} \, {\rm sgn}(\tau) .
\end{eqnarray}
Now $|g_{12}(\tau)|$ is exponentially small unless $\Delta|\tau| < 1$.
In particular, $g_{12}(\tau) \to -\delta(\tau)$ for $\Delta \to \infty$.
Moreover, for $B\ll \Delta$ with $B\equiv |{\bf B}|$, the magnetic impurity (${\bf S}$) 
dynamics will be slow on time scales of order $1/\Delta$. 
We may therefore approximate
the spin-spin correlators in Eq.~(\ref{I2}) by their respective equal-time expressions,
\begin{equation}\label{Q12}
\lim_{\tau_1 \to \tau_2} \langle {\cal T}_\tau {\cal Q}_{\sigma, \uparrow}(\tau_1) {\cal Q}_{-\sigma, \uparrow}(\tau_2) \rangle^{}_0 =
\frac{\sigma}{4} {\rm sgn}(\tau_1 - \tau_2) \langle S_+(\tau_1) \rangle^{}_0 .
\end{equation}
Inserting Eqs.~(\ref{g12}) and (\ref{Q12}) into the expression for the supercurrent in Eq.~(\ref{I2}), 
the time integrations can be carried out analytically. 

We obtain the CPR in the cotunneling regime as 
\begin{eqnarray}\label{I2xy}
I(\phi) &=& I_x \sin \phi + I_y \cos \phi ,\\ \nonumber
I_{x,y}&=& \frac{e\kappa^2 \Delta_p}{2\hbar} \frac{B_{x,y}}{B}  \tanh(\beta B),
\end{eqnarray}
with $\kappa$ in Eq.~(\ref{kappa}).  We note that while $I(\phi)$ is formally independent of $\Delta$, the value of 
$\Delta$ must be sufficiently large to justify the steps leading to Eq.~(\ref{I2xy}).
Remarkably, Eq.~(\ref{I2xy}) predicts anomalous supercurrents for
the S-QD-TS setup, i.e., a finite Josephson current for vanishing phase difference ($\phi = 0$) \cite{Buzdin2008,Zazunov2009,Brunetti2013}.
One can equivalently view this effect as a $\varphi_0$-shift in the CPR, 
$I(\phi)=I_c \sin(\phi+\varphi_0)$. An observation of this $\varphi_0$-junction behavior could  
then provide additional evidence for MBSs (see also Ref.~\cite{Schrade2017}), where
Eq.~(\ref{I2xy}) shows that the local magnetic field is required to have a finite $B_y$-component with $\hat e_z$ defining  
 the MBS spin polarization direction.  In particular,
if ${\bf B}$ is aligned with $\hat e_z$, 
the supercurrent in Eq.~(\ref{I2xy}) vanishes identically  since 
$s$-wave Cooper pairs cannot tunnel from the S lead into the TS wire in the absence of spin flips 
\cite{Zazunov2012}.  Otherwise, the CPR is $2\pi$-periodic and 
sensitive to the MBS through the peculiar dependence on the
relative orientation between the MBS spin polarization ($\hat e_z$) and the local Zeeman field ${\bf B}$ on the QD.   The fact that $B_y\ne 0$ (rather than $B_x\ne 0$) is necessary to have $\varphi_0\ne 0$ can be traced back to our choice of real-valued tunnel couplings. For tunable
tunnel phases, also the field direction where one has $\varphi_0=0$ will vary accordingly.

Noting that the anomalous Josephson effect has recently been  observed in S-QD-S devices \cite{Szombati2016}, 
 we expect that similar experimental techniques will allow to access the CPR (\ref{I2xy}).
We mention in passing that previous work has also pointed out that experiments 
employing QDs between N (instead of S) leads and TS wires can probe nonlocal effects due
to MBSs   
 \cite{Deng2016,Aguado2017,Flensberg2011,Sticlet2012,Prada2012,Prada2017,Hoffman2017}.
In our case, e.g., by variation of the field direction in the $xy$-plane, 
Eq.~(\ref{I2xy}) predicts a tunable anomalous supercurrent.
We conclude that in the cotunneling regime, the $\pi$-junction behavior of S-QD-S devices 
is replaced by the more exotic physics of $\varphi_0$-junctions in the S-QD-TS setting. 

\subsubsection{Mean-field approximation}

Next we present a mean-field analysis of the Hamiltonian (\ref{htot}) which allows us to go beyond the perturbative
cotunneling regime. For the corresponding S-QD-S case, see~Refs.~\cite{Vecino2003,Alvaro2012}.  
We note that a full solution of this interacting many-body problem requires a detailed numerical analysis using, e.g.,
the numerical renormalization group \cite{Choi2004,Karrasch2008} or quantum Monte Carlo simulations \cite{Siano2004,Luitz2012},
which is beyond the scope of the present work. 
We start by defining the GF of the QD, 
\begin{equation}
G_d(\tau) = - \langle {\cal T}_{\tau} \Psi_d(\tau) \Psi_d^{\dagger}(0) \rangle,\quad
\Psi_d^{\dagger} = (d^\dag_\up, d_\dw, d^\dag_\dw, -d_\up)^T.
\end{equation}
Note that this notation introduces double counting, which implies that only half of the levels are physically independent.
Of course, the results below take this issue into account.

With the above Nambu bi-spinor basis, the mean-field Hamiltonian has the $4\times 4$ matrix representation
\begin{eqnarray} \label{mfham} 
{\cal H}_{\rm MF} &=& \left(\begin{array}{cccc} \epsilon_\up & \Delta_d & \alpha_d & 0 \\
           \Delta_d^* & -\epsilon_\dw & 0 & \alpha_d \\
           \alpha_d^* & 0 &  \epsilon_\dw & \Delta_d \\
           0 & \alpha_d^* &\Delta_d^* &  - \epsilon_\up \end{array}\right),\\
\nonumber
\epsilon_\up &=& \epsilon_0  - B_z + U\langle n_\dw \rangle  , \quad 
\epsilon_\dw = \epsilon_0 + B_z + U   \langle n_\up \rangle , \\
\alpha_d &=&  B_x + iB_y - U \bra d^\dag_\dw d_\up \ket , \quad
\Delta_d = U \bra d_\dw d_\up  \ket.  \nonumber
\end{eqnarray}
The mean-field parameters appearing in Eq.~(\ref{mfham}) follow by solving the self-consistency equations 
\begin{eqnarray}\label{sceq}
&& \langle n_\up \rangle = \frac{1}{\beta} \sum_{\omega} G_{d,11}(\omega), \quad
\langle n_\dw  \rangle = \frac{1}{\beta} \sum_{\omega} G_{d,33}(\omega) ,\\
&& \bra d^\dag_\dw d_\up \ket  = \frac{1}{\beta} \sum_{\omega} G_{d,13}(\omega) , \nonumber \quad
\bra d_\dw d_\up \ket = \frac{1}{\beta} \sum_{\omega} G_{d,21}(\omega)  ,
\end{eqnarray}
where the mean-field approximation readily yields
\begin{equation}
G_d(\omega) = \left[i\omega- {\cal H}_{\rm MF} - \Sigma_{S}(\omega) - \Sigma_{TS}(\omega)\right]^{-1}.
\label{fullGF}
\end{equation}
The self-energies $\Sigma_{S/TS}(\omega)$ due to the coupling of the QD to the S/TS leads have the  matrix representation
\begin{equation}\label{sigma1}
\Sigma_{S} = \Gamma_S \left ( \begin{array}{cccc} g_{11} & -g_{12} & 0 & 0 \\
-g_{21} & g_{22} & 0 &0 \\ 0 & 0& g_{11}& -g_{12} \\
0 & 0 &-g_{21} & g_{22} \end{array}\right) 
\end{equation}
and
\begin{equation}\label{sigma2}
\Sigma_{\rm TS} = \Gamma_{TS} 
\left(\begin{array}{cccc} G_{11} & 0 & 0 & -G_{12}e^{-i\phi} \\
0 & 0 & 0 & 0 \\ 0 & 0 & 0 & 0 \\
-G_{21}e^{i\phi} & 0 & 0 & G_{22}\end{array}\right)
\end{equation}
with the hybridization parameters $\Gamma_{S/TS} = \lambda_{S/TS}^2$.  The bGFs $g(\omega)$ and $G(\omega)$
have been defined in Eqs.~(\ref{g}) and (\ref{G}), respectively.
Once a self-consistent solution to Eq.~(\ref{sceq}) has been determined, which in general requires numerics, the Josephson current 
is obtained from Eq.~(\ref{hatI}) as
\begin{equation}\label{Iphi}
I(\phi) = -\frac {e}{\hbar\beta} \sum_\omega \frac{\partial_\phi {\rm det} \left[ G_d^{-1}(\omega) \right]}{
{\rm det} \left[ G_d^{-1}(\omega) \right]}.
\end{equation}
In what follows, we study a setup with $\Delta_p=\Delta$ and consider the zero-temperature limit.

In order to compare our self-consistent mean-field results to the noninteracting case, let us briefly summarize analytical expressions for the $U=0$ ABS spectrum in the atomic limit 
defined by $\Gamma_{S,TS}\ll \Delta$. 
First we notice that at low energy scales, the self-energy $\Sigma=\Sigma_S+\Sigma_{TS}$, see Eqs.~(\ref{sigma1}) and (\ref{sigma2}), 
simplifies to 
\begin{equation}    \label{sigmaap}
\Sigma \simeq \left(\begin{array}{cccc}   \frac{2\Delta}{i\omega}\Gamma_{TS}  & -\Gamma_{S} & 0&  -\frac{2\Delta}{i\omega}\Gamma_{TS} e^{-i\phi} \\
-\Gamma_S &  0& 0 & 0 \\
0 & 0 &  0& -\Gamma_S \\
-\frac{2\Delta}{i\omega}\Gamma_{TS} e^{i\phi} & 0 &- \Gamma_S &  \frac{2\Delta}{i\omega}\Gamma_{TS} 
\end{array}\right). 
\end{equation}
The ABS spectrum of the S-QD-TS junction then follows by solving a determinantal equation, ${\rm det}\left[G_d^{-1}(\omega)\right]=0$.
One finds a zero-energy pole which is related to the MBS and results from the 
$1/\omega$ dependence of $\Sigma_{TS}(\omega)$.  In addition, we get finite-energy subgap poles for
\begin{equation}\label{analytical}
i\omega\equiv  E_A^{(\sigma_1=\pm ,\sigma_2=\pm)}= \sigma_1 \sqrt{\frac{b_0+\sigma_2\sqrt{b_0^2+4c_0}}{2}} ,
\end{equation}
with the notation
\begin{eqnarray}\nonumber
&& b_0 = \epsilon_\dw^2 + \epsilon_\up^2 + 4\Gamma_{TS}\Delta + 2\Gamma_S^2 + 2|\alpha_d|^2,   \\ \nonumber
&& c_0 = -4\Gamma_{TS}\Delta \left( \epsilon_\dw^2 +\Gamma_S^2 + |\alpha_d|^2 \right) - \epsilon_\up^2\epsilon_\dw^2 \\
&& \qquad - \left( |\alpha_d|^2 - \Gamma_S^2 \right) \left( |\alpha_d|^2 - \Gamma_S^2 - 2\epsilon_\up\epsilon_\dw \right) \nonumber \\
&&  \qquad +\ 8\Delta\Gamma_S\Gamma_{TS} {\rm Re}\left( \alpha_d e^{i\phi} \right)  \label{phidep} .
\end{eqnarray}

\begin{figure}
\includegraphics[width=1.1\columnwidth]{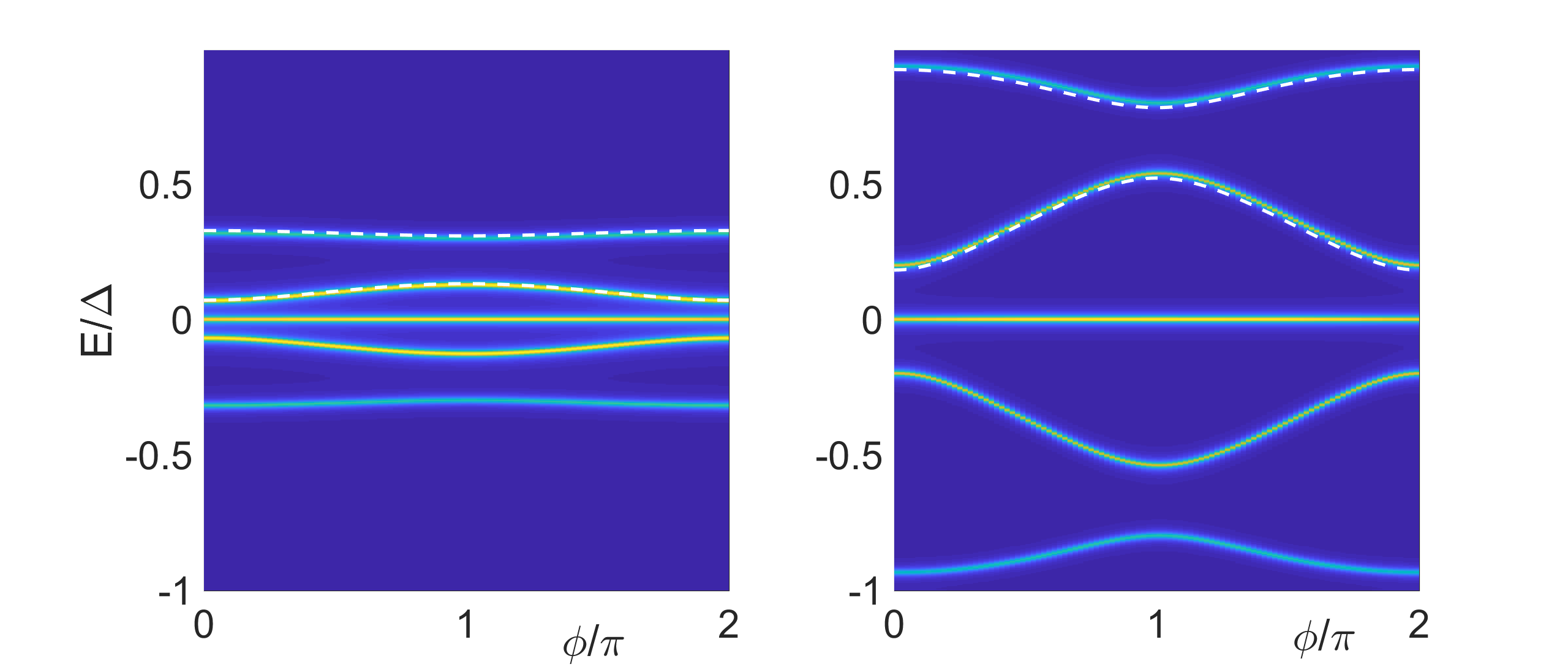}
\caption{Phase dependence of the subgap spectrum of an S-QD-TS junction in the noninteracting case, $U=0$.
The TS wire is modeled from the low-energy limit of a Kitaev chain, and we use the parameters
$B_y=0$, $B_x=B_z=B/\sqrt2$,  $\epsilon_0=0$, $\Delta_p=\Delta$, and $\Gamma_S=\Gamma_{TS}=\Gamma$.
From blue to yellow, the color code indicates increasing values of the spectral density.
The left (right) panel is for $\Gamma=0.045\Delta$ and $B=0.1\Delta$
($\Gamma=B=0.5\Delta$).  Solid curves were obtained by 
 numerical evaluation of Eq.~(\ref{Iphi}). Dashed curves give the analytical prediction (\ref{analytical}). 
In the right panel, the energies resulting from Eq.~(\ref{analytical}) have been rescaled by the factor $1 + \Gamma/\Delta$.}
\label{fig2}
\end{figure}

In Fig.~\ref{fig2}, numerically exact results for the $U=0$ ABS spectrum are compared to 
the analytical prediction (\ref{analytical}).   We first notice that, as expected, Eq.~(\ref{analytical}) accurately fits the
numerical results in the atomic limit, see the left panel in Fig.~\ref{fig2}.  Deviations can be observed for
larger values of $\Gamma_{S,TS}/\Delta$. However, as shown in the right panel of Fig.~\ref{fig2},  rather
good agreement is again obtained by rescaling Eq.~(\ref{analytical}) with a constant factor of the order of  $(1 + \Gamma_{S,TS}/\Delta)$.
For finite $B_y$, we find (data not shown) that the phase-dependent ABS spectrum is shifted with respect to $\phi=0$.  In fact, since 
the phase dependence of the subgap states comes from the term Re$(\alpha_d e^{i\phi})$ in the atomic limit, 
see Eqs.~(\ref{mfham}) and (\ref{phidep}),  $B_y$ can be fully accounted for in this limit by simply shifting $\phi\to \phi+\varphi_0$. 
We thereby recover the $\varphi_0$-junction behavior discussed before for the cotunneling regime, see Eq.~(\ref{I2xy}).

\begin{figure}
\includegraphics[width=1.1\columnwidth]{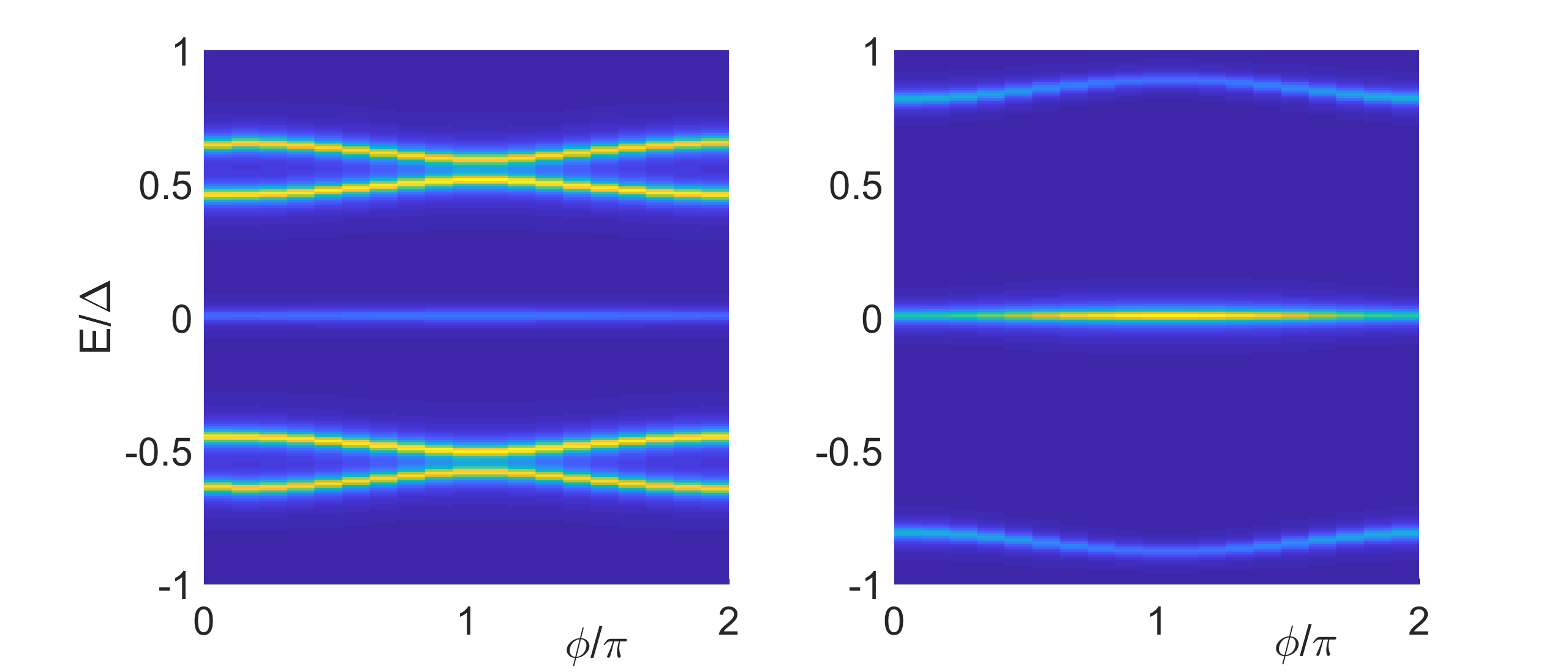}
\caption{Phase-dependent ABS spectrum from mean-field theory for S-QD-TS junctions as in Fig.~\ref{fig2} but with $U>0$ and $\epsilon_0=-U/2$. 
We put $\Delta_p=\Delta$, $B_y=0$, and $\Gamma_{S}=\Gamma_{TS}=\Gamma$.   
The color code is as in Fig.~\ref{fig2}. The left panel is for $U=\Delta$, $\Gamma=0.5\Delta$, and $B_x=B_z=B/\sqrt2$ with $B=0.5\Delta$ [cf.~the right panel
of Fig.~\ref{fig2}].
The right panel is for $U=10\Delta$, $\Gamma= 4.5\Delta$, $B_x=15\Delta$, and $B_z=0$.  }
\label{fig3}
\end{figure}

We next turn to self-consistent mean-field results for the phase-dependent ABS spectrum at finite $U$. 
Figure~\ref{fig3} shows the spectrum for the electron-hole symmetric case $\epsilon_0=-U/2$, with other
parameters as in the right panel of Fig.~\ref{fig2}. 
For moderate interaction strength, e.g., taking $U= \Delta$ (left panel), we find that compared to the $U=0$ case in Fig.~\ref{fig2}, 
interactions push together pairs of Andreev bands, e.g., the pair corresponding to $E_A^{(+,\pm)}$ in Eq.~(\ref{Iphi}). 
 On the other hand, for stronger interactions, e.g., $U =10\Delta$ (right panel),
 the outer ABSs leak into the continuum spectrum and only the inner Andreev states remain inside the superconducting gap.
 The ABS spectrum shown in Fig.~\ref{fig3} is similar to what is observed in mean-field calculations for S-QD-S systems with broken
 spin symmetry and in the magnetic regime of the QD, where one finds up to four ABSs for $U<\Delta$ while 
 the outer ABSs merge with the continuum for $U> \Delta$ \cite{Alvaro2012}. 
 Interestingly, the inner ABS contribution to the free energy for $U=10\Delta$  is 
 minimal for $\phi=\pi$, see right panel of Fig.~\ref{fig3},
 and we therefore expect $\pi$-junction 
 behavior for $B_y=0$ also in the regime with $U\gg \Delta$ and $B\gg \Delta$.
 We notice, however, that changing the sign of $B_x$ would result in zero
 junction behavior. 
 We interpret the inner ABSs for $U\gg \Delta$ as Shiba states with the phase dependence generated by the coupling to the MBS.  Without the latter coupling, 
 the Shiba state has $\phi$-independent energy slightly below $\Delta$ determined by the scattering phase shift difference between both spin polarizations \cite{Balatsky2006}.

\begin{figure}
\includegraphics[width=0.95\columnwidth]{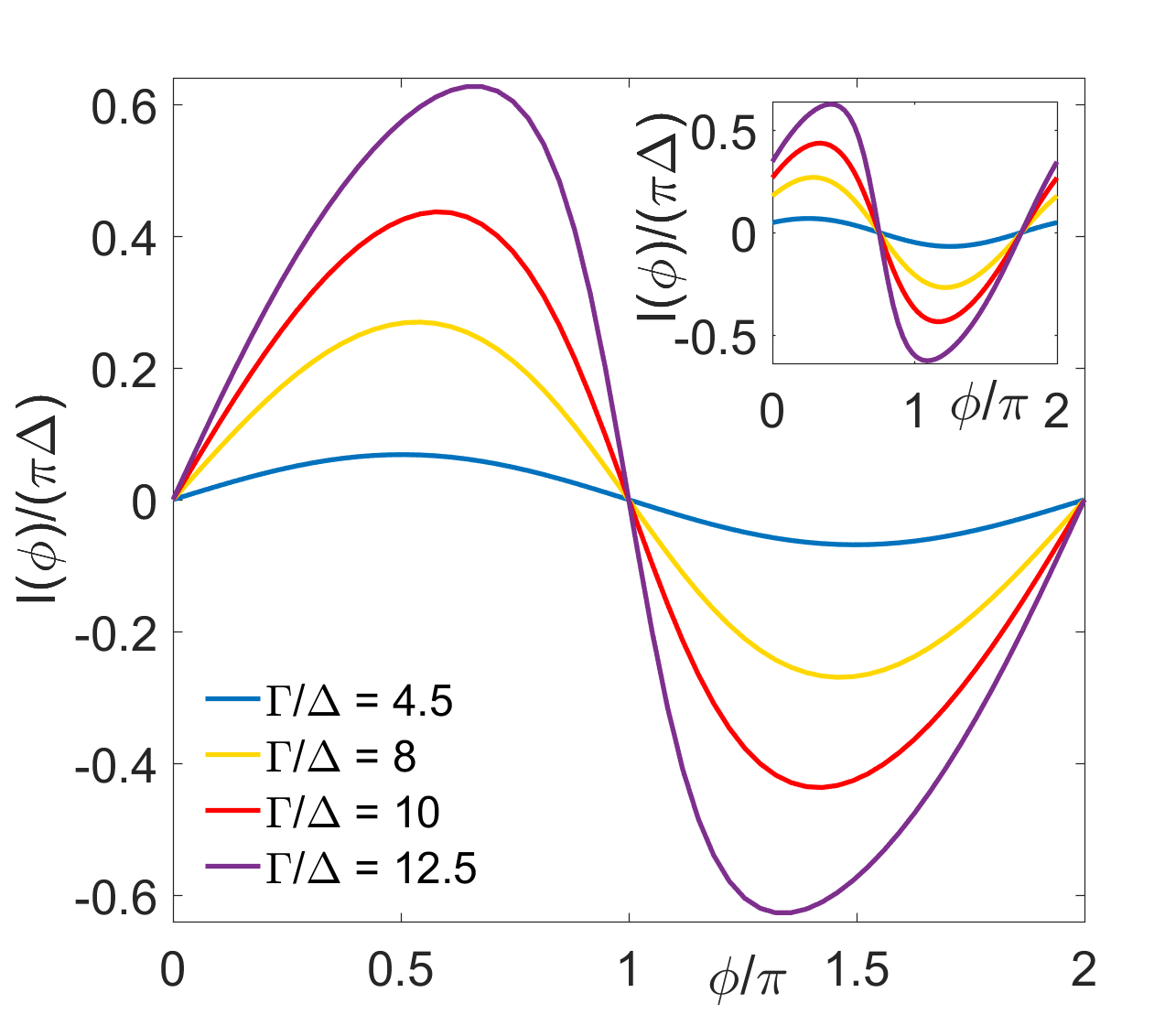}
\caption{Main panel: Mean-field results for the CPR of S-QD-TS junctions with different $\Gamma/\Delta$ values, where we assume $\Delta_p=\Delta$, $U=10\Delta$, 
$\epsilon_0=-U/2$, $\Gamma_{S}=\Gamma_{TS}=\Gamma$, $B=15\Delta$, and $B_z=0$.
Main panel: For $B_x=-B$ and $B_y=0$, we find $\pi$-junction behavior.
Inset: Same but for $B_y=-B_x=B/\sqrt2$, where $\varphi_0$-junction behavior occurs. }
\label{fig4}
\end{figure}

 As illustrated in Fig.~\ref{fig4}, the CPR computed numerically from Eq.~(\ref{Iphi}) for different values of $\Gamma_{S,TS}/\Delta$, where
 $B_x$ has been inverted with respect to its value in Fig.~\ref{fig3}, results 
in zero junction behavior for $B_y=0$.  
 This behavior is expected from Eq.~(\ref{I2xy}) in the cotunneling regime, and Fig.~\ref{fig4} shows that it also 
 persists for $\Gamma_{S,TS}\gg \Delta$.  
In contrast to Eq.~(\ref{I2xy}), however, the CPR for $\Gamma_{S,TS} \gg  \Delta$ differs from a purely 
sinusoidal behavior, see Fig.~\ref{fig4}.  Moreover,  for $B_y\ne 0$, we 
again encounter $\varphi_0$-junction behavior, cf.~the inset of Fig.~\ref{fig4}, in accordance with the perturbative result in Eq.~(\ref{I2xy}).
Our mean-field results suggest that $\varphi_0$-junction behavior is very robust and extends also into other parameter regimes as long as the
condition $B_y\ne 0$ is met.

\begin{figure}
\includegraphics[width=0.95\columnwidth]{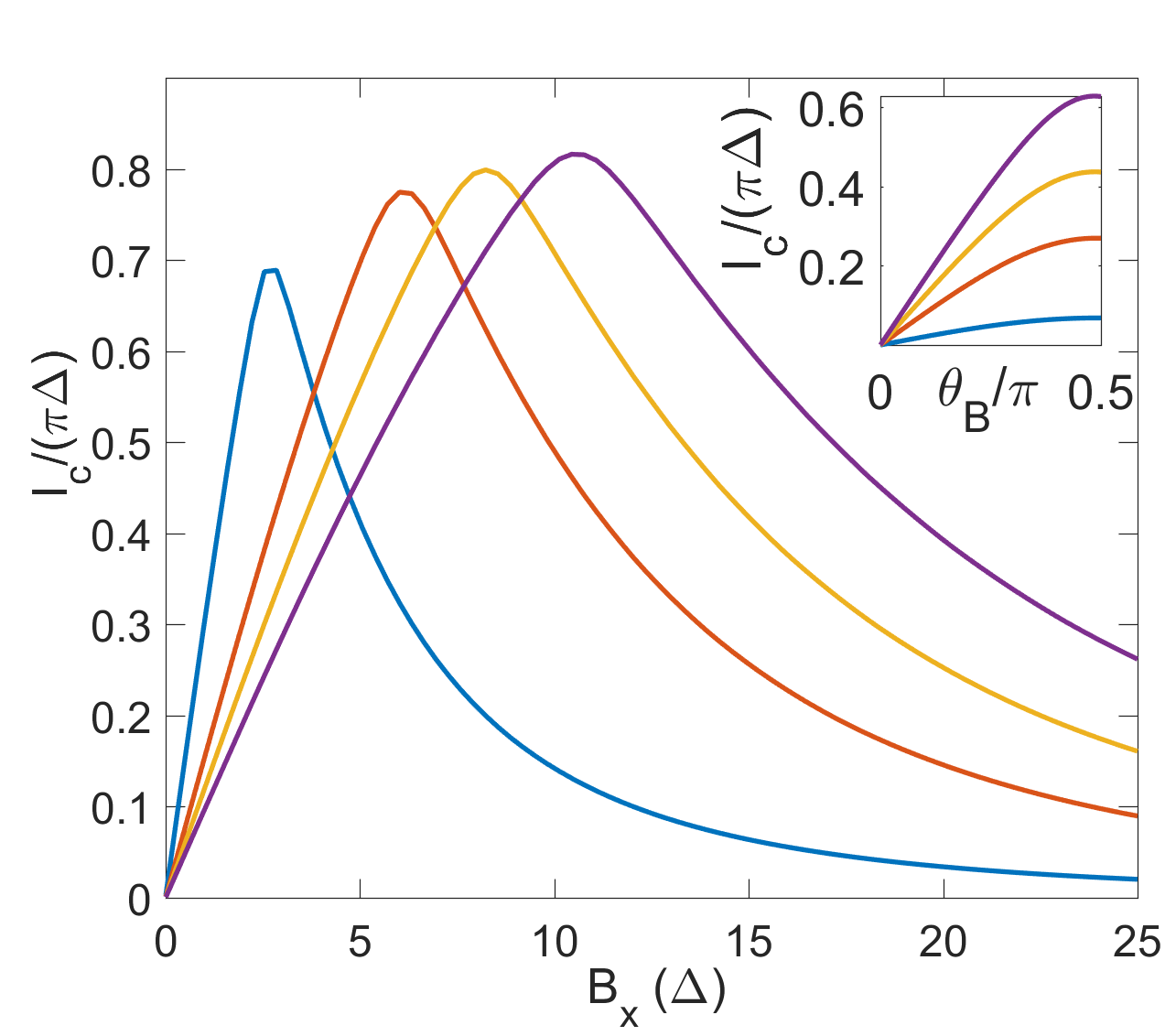}
\caption{Main panel: Mean-field results for the critical current $I_c$ vs local magnetic field scale $B_x$ in S-QD-TS junctions.
Parameters are as in the main panel of Fig.~\ref{fig4}, i.e.,
$U=10\Delta$, $\epsilon_0=-U/2$,  and $B_{y,z}=0$.  From left to right,
different curves are for $\Gamma/\Delta=4.5, 8, 10$ and 12.5.  
Inset: $I_c$ vs angle $\theta_B$, where ${\bf B}=B (\sin\theta_B,0,\cos\theta_B)$ with
$B=15\Delta$. }
\label{fig5}
\end{figure}

Next, Fig.~\ref{fig5} shows mean-field results for the critical current, $I_c={\rm max}_\phi |I(\phi)|$, as function of 
the local magnetic field $B_x$ and otherwise the same parameters as in Fig.~\ref{fig4}.  The main panel in Fig.~\ref{fig5} shows that
$I_c$ increases linearly with $B_x$ for small $B_x < \Delta$, then exhibits 
a maximum around $B_x \approx \Gamma$, and subsequently decreases again to small values 
for $B_x \gg {\rm max}\{\Gamma_{S,TS}, \Delta\}$.
On the other hand, for a fixed absolute value $B$ of the magnetic field and $B_y=0$, the critical current also exhibits a
maximum as a function of the angle $\theta_B$ between ${\bf B}$ and the MBS spin polarization axis ($\hat e_z$). This effect is illustrated in the inset of
 Fig.~\ref{fig5}. As expected, the Josephson current vanishes for $\theta_B \to 0$, where the supercurrent blockade argument of Ref.~\cite{Zazunov2012} 
implies $I_c=0$, and reaches its maximal for $\theta_B = \pi/2$.  

\subsection{Spinful nanowire model for the TS} 

\subsubsection{Model}

Before turning to the S-TS-S setup in Fig.~\ref{fig1}(b), we address the 
question of how the above  results for S-QD-TS junctions  change when using the spinful nanowire model of Refs.~\cite{Lutchyn2010,Oreg2010} instead of the
low-energy limit of a Kitaev chain, see Eq.~(\ref{G}).
In fact, we will first describe the Josephson current for the elementary case of an S-TS 
junction using the spinful nanowire model.  Surprisingly,
to the best of our knowledge, this case has not yet been addressed in the literature. 

In spatially discretized form, the spinful nanowire model for TS wires 
reads \cite{Lutchyn2010,Oreg2010,Zazunov2017}
\begin{eqnarray}\label{oregmodel}
H_{\rm TS} &=& \frac12 \sum_j \left[\psi_j^{\dagger} \hat{h} \psi_j^{} + 
\left( \psi_j^{\dagger} \hat{t} \psi_{j+1}^{}  + \mbox{h.c.}\right)\right],\\
\nonumber 
\hat{h} &=& (2t-\mu)\tau_z \sigma_0 + V_x \tau_0\sigma_x + 
\Delta_p \tau_x\sigma_0,\\ \nonumber
\hat{t} &=& -t\tau_z\sigma_0 + i\alpha\tau_z\sigma_z,
\end{eqnarray}
where the lattice fermion operators $c_{j\sigma}$ for 
given site $j$ with spin polarizations $\sigma=\uparrow,\downarrow$ 
are combined to the four-spinor operator 
$\psi_j= \left(c_{j\uparrow}^{}, 
c_{j\downarrow}^{}, c^{\dagger}_{j\downarrow}, -c^{\dagger}_{j\uparrow} \right)^T$. 
The Pauli matrices $\tau_{x,y,z}$ (and unity $\tau_0$) again act in Nambu space, while 
Pauli matrices $\sigma_{x,y,z}$ and $\sigma_0$ refer to spin.
In the figures shown below, we choose the model parameters in Eq.~(\ref{oregmodel})
as  discussed in Ref.~\cite{Zazunov2017}. The lattice spacing is set to $a=10$~nm,
which results in a nearest-neighbor hopping $t = \hbar^2/(2m^*a^2)= 20$~meV  
and the spin-orbit coupling strength $\alpha=4$~meV for InAs nanowires.  
The proximity-induce pairing gap is again denoted by $\Delta_p$, the chemical potential is $\mu$, and
 the bulk Zeeman energy scale $V_x$ is determined by a magnetic field applied along the wire.  
Under the condition
\begin{equation}\label{topcrit}
V_x>V_x^c=\sqrt{\mu^2+\Delta_p^2},
\end{equation}  
the topologically nontrivial phase is realized \cite{Lutchyn2010,Oreg2010}.
As we discuss below, the physics of the S-QD-TS junction  sensitively depends on both the bulk Zeeman field $V_x$ 
and on the local magnetic field ${\bf B}$ acting on the QD, where one can either 
identify both magnetic fields or treat ${\bf B}$ as independent field.  In any case,
the bGF $\tilde G(\omega)$ for the model in Eq.~(\ref{oregmodel}), 
which now replaces the Kitaev chain result $G(\omega)$ in Eq.~(\ref{G}),
needs to be computed numerically.  
The bGF  $\tilde G$ has been described in detail in Ref.~\cite{Zazunov2017}, where also a straightforward numerical 
scheme for calculating $\tilde G(\omega)$ has been devised.   
With the replacement $G\to \tilde G$, 
we can then take over the expressions for the Josephson current discussed before.
Below we study these expressions in the zero-temperature limit.

\subsubsection{S-TS junction}

Let us first address the CPR for the S-TS junction case.  
The Josephson current can be computed using the bGF expression for
 tunnel junctions in Ref.~\cite{Zazunov2016}, which is a simplified version of the 
above expressions for the S-QD-TS case.
The spin-conserving tunnel coupling $\lambda$ defines a transmission probability (transparency)
${\cal T}$ of the normal junction \cite{Zazunov2016,Zazunov2017}. Close to the topological transition, the transparency is well approximated by
\begin{equation}\label{transp}
{\cal T}= \frac{4(\lambda/t)^2}{[1+(\lambda/t)^2]^2},
\end{equation}
where $t= 20$~meV is the hopping parameter in Eq.~(\ref{oregmodel}).
We then study the CPR and the resulting critical current $I_c$ as a function of ${\cal T}$ for both the topologically 
trivial ($V_x<V_x^c$) and the nontrivial ($V_x>V_x^c$) regime, see Eq.~(\ref{topcrit}).

\begin{figure}
\includegraphics[width=0.95\columnwidth]{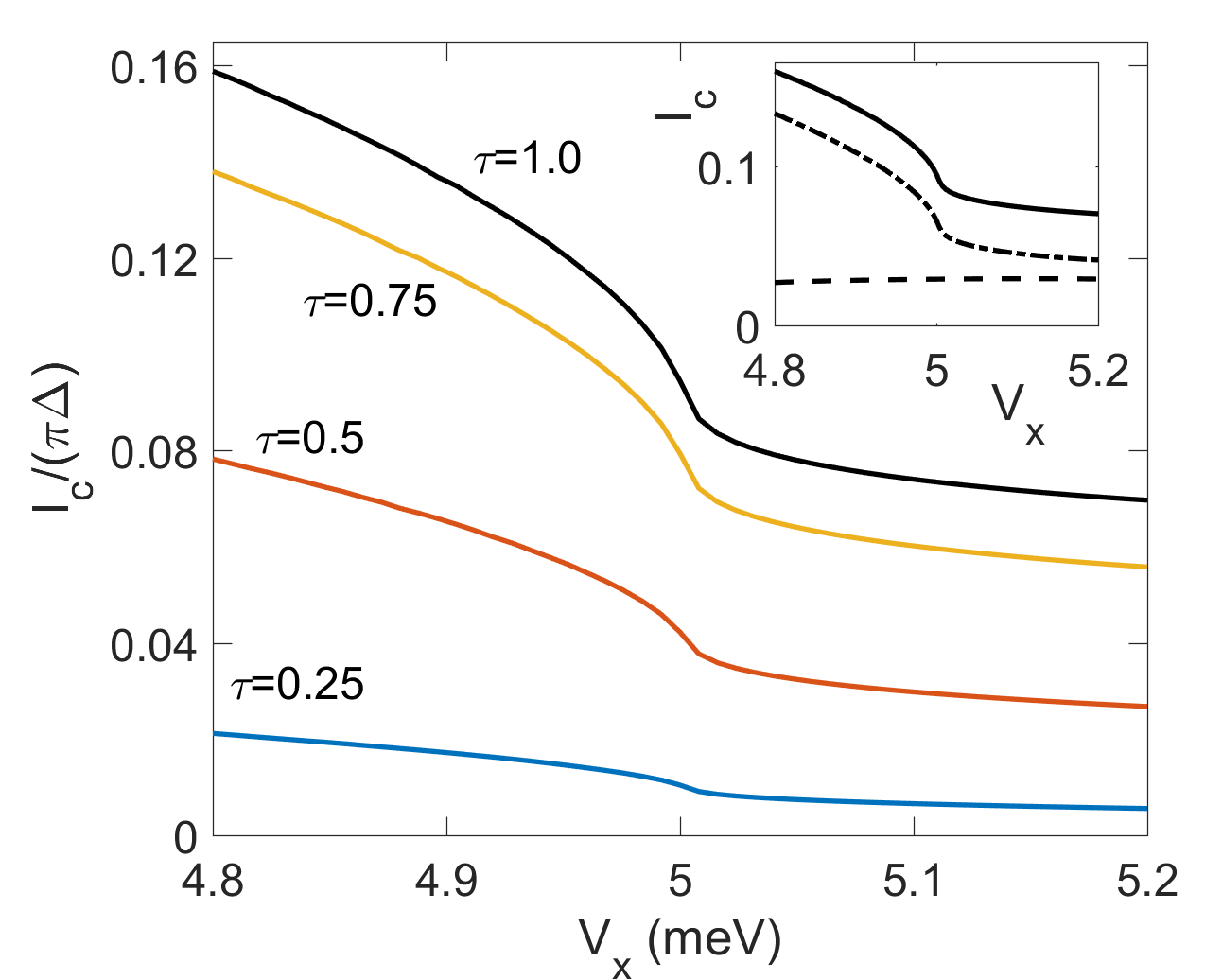}
\caption{
Main panel: Critical current $I_c$ vs Zeeman energy $V_x$ for an S-TS junction 
using the spinful TS nanowire model (\ref{oregmodel})
for $\Delta_p=\Delta=0.2$~meV, $\mu=5$~meV, and different transparencies ${\cal T}$ calculated from Eq.~(\ref{transp}). 
All  other parameters are specified in the main text.
Inset: Decomposition of $I_c$ for ${\cal T}=1$ into ABS (dotted-dashed) and continuum (dashed) contributions. 
}
\label{fig6}
\end{figure}

In Fig.~\ref{fig6}, we show the $V_x$ dependence of the critical current $I_c$ for the symmetric case $\Delta=\Delta_p$. In particular, it is of interest to determine
how $I_c$ changes as one moves through the phase transition in Eq.~(\ref{topcrit}). First, we observe that $I_c$ is strongly suppressed in the topological phase in 
comparison to the topologically trivial phase. In fact, $I_c$ slowly decreases as 
one moves into the deep topological phase by increasing $V_x$. This observation is in accordance with the expected supercurrent 
blockade in the deep topological limit \cite{Zazunov2012}: $I_c=0$ for the corresponding Kitaev chain case
since $p$-wave pairing correlations on the TS side are incompatible with $s$-wave correlations on the S side.   
 However, a residual finite supercurrent can be observed even for rather large values of $V_x$.
We attribute this effect to the remaining $s$-wave pairing correlations contained in the spinful nanowire model (\ref{oregmodel}). 
Second, Fig.~\ref{fig6} shows kink-like features in the $I_c(V_x)$ curve near the topological transition, $V_x\approx V_x^c$.
The inset of Fig.~\ref{fig6} demonstrates that this feature comes from a rapid decrease of the ABS contribution while the continuum contribution remains smooth. 
This observation suggests that continuum contributions in this setup mainly originate from $s$-wave pairing correlations which are not particularly sensitive to the topological transition.

\begin{figure}
\includegraphics[width=0.95\columnwidth]{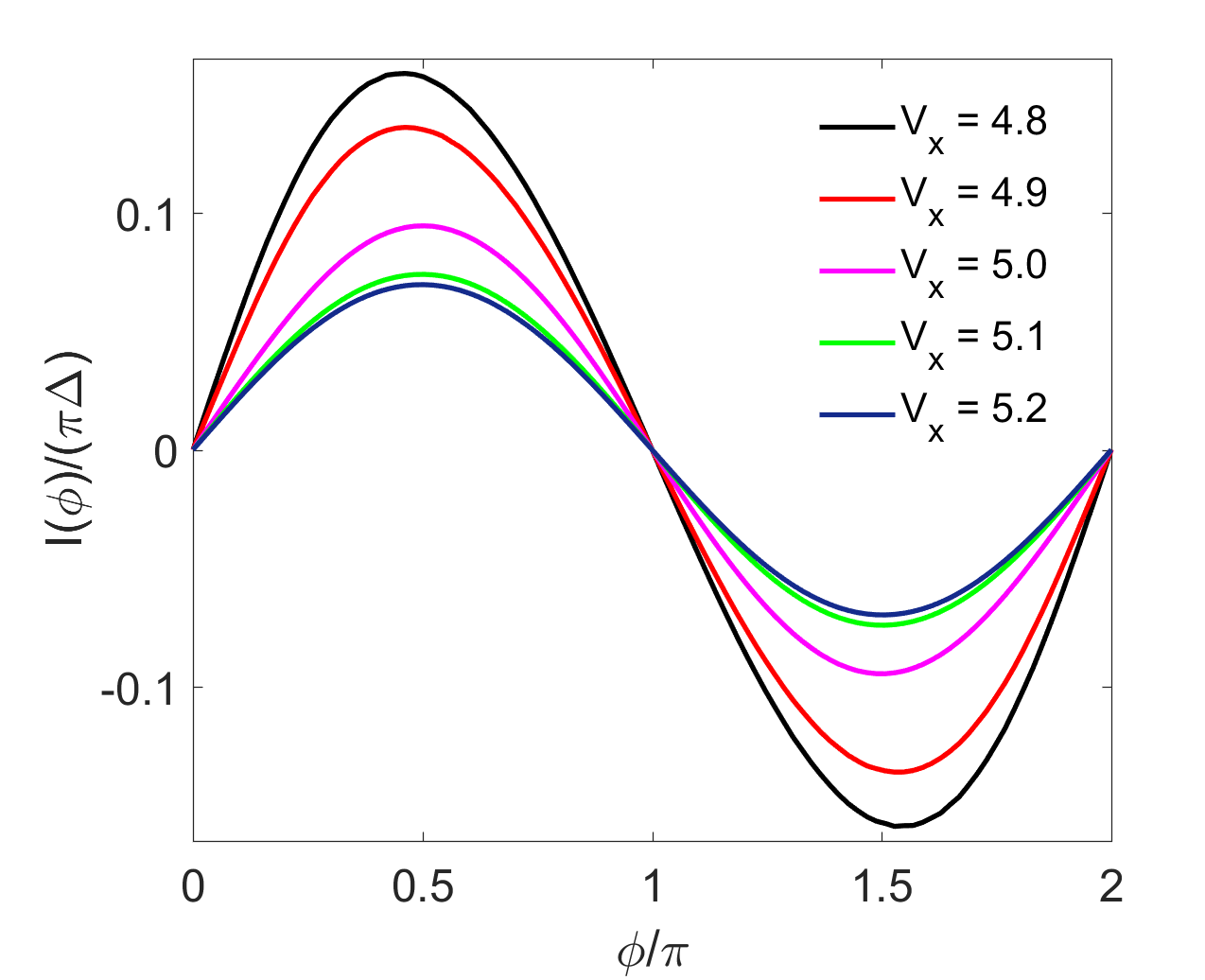}
\caption{
CPR for the S-TS junction with ${\cal T}=1$ in Fig.~\ref{fig6}, for different bulk Zeeman fields $V_x$ (in meV)
near the critical value $V_x^c=5.004$~meV.
}
\label{fig7}
\end{figure}

In Fig.~\ref{fig7}, we show the CPR for the S-TS junction with ${\cal T}=1$ in Fig.~\ref{fig6}, where different curves correspond to 
different Zeeman couplings $V_x$ near the critical value. 
We find that in many parameter regions, in particular for ${\cal T}<1$, the CPR is to high accuracy given by
a conventional $2\pi$-periodic Josephson relation, $I(\phi) = I_c \sin\phi$.
In the topologically trivial phase, small deviations from the sinusoidal law can be detected, but once one enters 
the topological phase, these deviations become extremely small.  

\subsubsection{S-QD-TS junction with spinful TS wire: Mean-field theory}

Apart from providing a direct link to experimental control parameters, another 
advantage of using the spinful nanowire model of Refs.~\cite{Lutchyn2010,Oreg2010}
for modeling the TS wire is that the angle between 
the local Zeeman field ${\bf B}$ and the MBS spin polarization does not have to be introduced as 
phenomenological parameter but instead results from the calculation \cite{Zazunov2017}. It is thus interesting to study 
the Josephson current in S-QD-TS junctions where the TS wire is described by the spinful nanowire model. 
For this purpose, we now  revisit the mean-field scheme for 
S-QD-TS junctions using the bGF $\tilde G(\omega)$ for the spinful nanowire model (\ref{oregmodel}).  In particular, with the 
replacement $G\to \tilde G$, we solve the self-consistency equations (\ref{sceq}) and thereby obtain
the mean-field parameters in Eq.~(\ref{mfham}).  The resulting QD GF, $G_d(\omega)$ in Eq.~(\ref{fullGF}), then determines the 
Josephson current in Eq.~(\ref{Iphi}).
Below we present self-consistent mean-field results obtained from this scheme.  In view of the huge parameter space of this problem, 
we here only discuss a few key observations. A full discussion of the phase diagram and the corresponding physics will be given elsewhere.

\begin{figure}
\includegraphics[width=0.95\columnwidth]{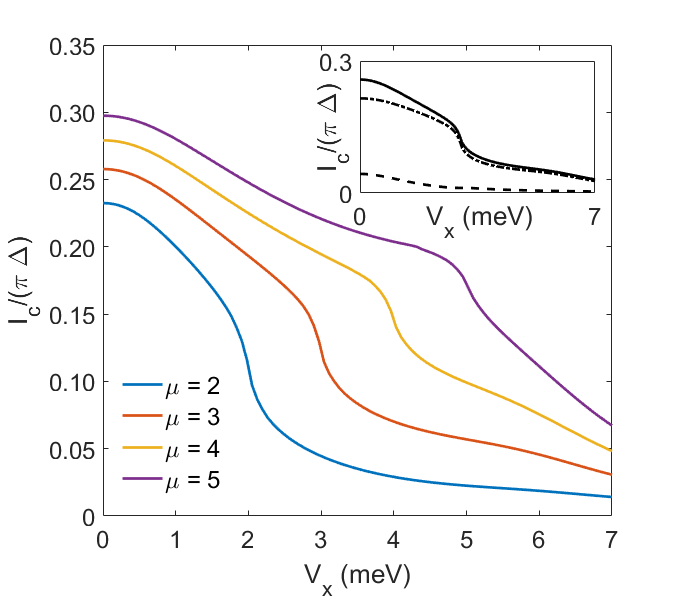}
\caption{Main panel: Critical current $I_c$ vs Zeeman energy $V_x$ for S-QD-TS junctions 
from mean-field theory using the spinful TS nanowire model (\ref{oregmodel}). Results are
shown for several values of the chemical potential $\mu$ (in meV), where we assume
$U=10\Delta$, $\epsilon_0 = -U/2$, $\Delta_p=\Delta=0.2$~meV, 
$\Gamma_S=2\Gamma_{TS}=9\Delta$, and ${\bf B}=(V_x,0,0)$.  
Inset: Detailed view of the transition region $V_x\approx V_x^c$ for $\mu=4$ meV, including
a decomposition of $I_c$ into the ABS (dotted-dashed) and the continuum
(dashed) contribution.  } 
\label{fig8}
\end{figure}

The main panel of Fig.~\ref{fig8} shows the critical current $I_c$ vs the bulk Zeeman
energy $V_x$ for several values of the chemical potential $\mu$, where the respective critical value  $V_x^c$ in Eq.~(\ref{topcrit}) for the topological phase
transition also changes with $\mu$.  The results in Fig.~\ref{fig8} assume that the local magnetic field ${\bf B}$ acting on the QD 
coincides with the bulk Zeeman field $V_x$ in the TS wire, i.e., ${\bf B}=( V_x,0,0)$.
For the rather large values of $\Gamma_{S,TS}$ taken in Fig.~\ref{fig8}, the  $I_c$ vs $V_x$ curves again exhibit a kink-like feature near the 
topological transition, $V_x\approx V_x^c$.  This behavior is very similar to what happens in S-TS junctions with large transparency ${\cal T}$, cf.~Fig.~\ref{fig6}.
As demonstrated in the inset of Fig.~\ref{fig8}, the physical reason for the kink feature can be traced back to a sudden drop of the 
ABS contribution to $I_c$ when entering the topological phase $V_x>V_x^c$.  In the latter phase, 
$I_c$ becomes strongly suppressed, 
in close analogy to the S-TS junction case shown in Fig.~\ref{fig6}.

\begin{figure}
\includegraphics[width=0.95\columnwidth]{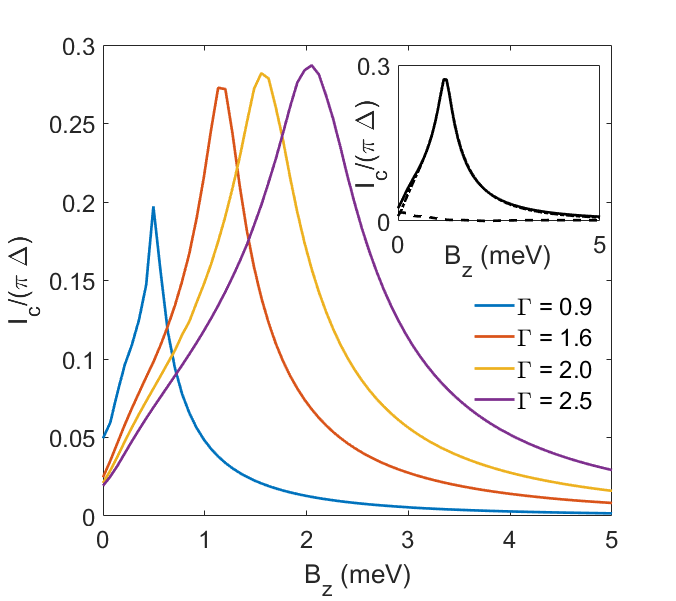}
\caption{Main panel: Mean-field results for $I_c$  vs $B_z$ in S-QD-TS junctions for several 
values of $\Gamma_S=\Gamma_{TS}=\Gamma$ (in meV) and $\mu=4$~meV.  The bulk Zeeman field $V_x=5$~meV along $\hat e_x$ (where $V_x>V_x^c$ for our parameters) 
is applied to the spinful TS wire, 
while the QD is subject to the local magnetic field ${\bf B}=B_z\hat e_z$.
All other parameters are as in Fig.~\ref{fig8}. 
Inset: Decomposition of $I_c$ into ABS (dotted-dashed) and continuum (dashed) 
contributions for $\Gamma=1.6$~meV. }
\label{fig9}
\end{figure}

In Fig.~\ref{fig8}, both the QD and the TS wire were
subject to the same magnetic Zeeman field.  If the direction and/or the size of the local magnetic field ${\bf B}$ applied to the QD 
can be varied independently from the bulk magnetic field $V_x\hat e_x$ applied to the TS wire, one can 
arrive at rather different conclusions. To illustrate this statement, Fig.~\ref{fig9} shows the $I_c$ vs $B_z$ dependence for  
${\bf B}=(0,0,B_z)$ perpendicular to the bulk field, with $V_x>V_x^c$ such that the TS wire
is in the topological phase.   In this case, Fig.~\ref{fig9} shows that
$I_c$ exhibits a maximum close to $B_z=0$. This behavior is reminiscent
of what we observed above in Fig.~\ref{fig5}, using the low-energy limit of a Kitaev chain for the bGF of the TS wire.
Remarkably, the critical current can  here reach values close to the unitary limit, $I_c\sim e\Delta/\hbar$. 
We note that since $B_z$ does not drive a phase transition, no kink-like features appear
for the $I_c(B_z)$ curves shown in Fig.~\ref{fig9}.
Finally, the inset of Fig.~\ref{fig9} shows that for ${\bf B}$ perpendicular to $V_x\hat e_x$,
where $V_x>V_x^c$ for the parameters chosen in Fig.~\ref{fig9},
 above-gap quasiparticles again provide a more significant contribution to $I_c$ than Andreev states.

\subsection{S-TS-S junctions: Switching the parity of a superconducting atomic contact}

\subsubsection{Model}

We now proceed to the three-terminal S-TS-S setup shown in Fig.~\ref{fig1}(b).
The CPR found in the related TS-S-TS trijunction case has been discussed in detail in Ref.~\cite{Zazunov2017}, see also Ref.~\cite{Sen2017}. Among other findings, a 
main conclusion of Ref.~\cite{Zazunov2017} for the TS-S-TS 
geometry was that the CPR can reveal information about the spin canting angle between the 
MBS spin polarization axes in both TS wires.
In what follows, we study the superficially similar yet rather different case of an
S-TS-S junction. Throughout this section, we model the TS wire via the low-energy theory of a spinless Kitaev chain, where the  bGF $G(\omega)$ in Eq.~(\ref{G}) applies. 

One can view the setup in Fig.~\ref{fig1}(b) as a conventional 
superconducting atomic contact (SAC) with a TS wire tunnel-coupled to 
the S-S junction.  Over the past few years, impressive experimental progress  \cite{Brethreau2013a,Brethreau2013b,Janvier2015} has demonstrated that the 
ABS level system in a SAC \cite{Zazunov2003} 
can be accurately probed and manipulated by coherent or incoherent microwave 
spectroscopy techniques.  
We show below that an additional TS wire, cf.~Fig.~\ref{fig1}(b), 
acts as tunable parity switch on the many-body ABS levels of the SAC.   
As we have discussed above, the supercurrent flowing directly between a given S lead and the 
TS wire is expected to be strongly suppressed. 
 However, through the  hybridization with the MBS, Andreev level 
 configurations with even and odd fermion parity are connected. 
 This effect has profound and potentially useful consequences for Andreev spectroscopy.

An alternative view of the setup in Fig.~\ref{fig1}(b) is to imagine an S-TS junction, where S1 plays the role 
of the S lead and the spinful TS wire is effectively composed from a spinless (Kitaev) TS wire and the S2 superconductor.
The $p$- and $s$-wave pairing correlations in the spinful TS wire are thereby spatially separated.
Since the $s$- and $p$-wave bands represent normal modes, the do not directly coupled to each other in this scenario, i.e., we have to put $\lambda_2=0$. 
We discuss this analogy in more detail later on. 

We consider a conventional single-channel SAC (gap $\Delta$) coupled via a point contact to a 
TS wire (gap $\Delta_{p}$), cf.~Fig.~\ref{fig1}(b).  
The superconducting phase difference across the SAC is denoted by
$\phi = \phi_1 - \phi_2$, where $\phi_j$ is the phase difference between  
the respective S arm ($j=1,2$) and the TS wire.  
In practice, the SAC can be embedded into a superconducting ring 
 for magnetic flux tuning of $\phi$.  
To allow for analytical progress, we here assume that $\Delta_p$ is so large 
that continuum quasiparticle excitations in the TS wire can be neglected. 
In that case, only the MBS at the junction has to be kept when modeling the
TS wire. However, we will also hint at how one can treat the general case.

For the two S leads,  boundary fermion fields are contained in Nambu spinors as in Eq.~(\ref{g}),
\begin{equation}\label{PsiSj}
\Psi_{S,j=1,2} = \left( \begin{array}{c} \psi_{j,\uparrow}\\  \ \psi_{j,\downarrow}^\dagger \end{array} \right),
 \end{equation}
where their bGF follows with the Nambu matrix $g(\omega)$ in Eq.~(\ref{g}) as
\begin{equation}\label{ggg}
g^{-1}_j(\omega)=g^{-1}(\omega) + b_j \tau_0.
\end{equation}
We again use Pauli matrices $\tau_{x,y,z}$ and unity $\tau_0$ in Nambu space.
The dimensionless parameters $b_{1,2}$ describe 
the Zeeman field component along the MBS spin polarization axis, see below. 
Since above-gap quasiparticles in the TS wire are neglected here,
the TS wire is represented by the Majorana 
operator $\gamma = \gamma^\dagger$, with $\gamma^2=1/2$, which anticommutes with all other fermions.
We may represent $\gamma$ by an auxiliary fermion $f_\uparrow$, where the index reminds us that the MBS
 spin polarization points along $\hat e_z$,
\begin{equation}\label{fdef}
\gamma= (f_\uparrow+f_\uparrow^\dagger)/\sqrt2.
\end{equation}
The other Majorana mode $\gamma'= -i(f_\uparrow-f_\uparrow^\dagger)/\sqrt2$,
which is localized at the opposite end of the TS wire, is assumed to have negligible hybridization
with the $\Psi_{S,j}$ spinors and with $\gamma$.
Writing the Euclidean action as $S = S_0 + S_{\rm tun}$, we have an uncoupled 
action contribution,
\begin{eqnarray}\nonumber
S_0  &=& \sum_{j=1,2}\int_0^\beta d \tau d \tau'  \bar \Psi_{S,j}(\tau) g^{-1}_j(\tau - \tau') \Psi_{S,j}(\tau')  + \\ &&\quad +
\frac12 \int_0^\beta d \tau \ \gamma(\tau) \partial_\tau \gamma(\tau).
\end{eqnarray}
The leads are connected by a time-local tunnel action corresponding to the tunnel Hamiltonian
\begin{eqnarray}\label{St}
H_{\rm tun} &=& t_0\left( \Psi^\dagger_{S,1} \tau_z e^{i\tau_z\phi/2} 
 \Psi_{S,2} + {\rm h.c.} \right) + \\
\nonumber && \quad + \sum_{j=1,2} \frac{\lambda_j}{\sqrt{2}} 
\left( \psi_{j,\uparrow}^\dagger e^{i \phi_j / 2} - {\rm h.c.} \right) \gamma .
\end{eqnarray}
Without loss of generality, we assume that the tunnel amplitudes $t_0$ and $\lambda_{1,2}$,
see Fig.~\ref{fig1}(b), are real-valued and that they include density-of-state factors again.
The parameter $t_0$ (with $0\le t_0\le 1$) determines the transparency ${\cal T}$ of 
the SAC in the normal-conducting state \cite{Nazarov2009}, cf.~Eq.~(\ref{transp}),
\begin{equation}\label{tau0}
{\cal T} = \frac{4 t_0^2}{ (1 + t_0^2)^2} .
\end{equation}
Note that in Eq.~(\ref{St}) we have again assumed spin-conserving tunneling, where
only spin-$\uparrow$ fermions in the SAC are tunnel-coupled to the
Majorana fermion $\gamma$, cf.~Eq.~(\ref{ht}).

At this stage, it is convenient to trace out the $\Psi_{S,2}$ spinor field.  
As a result, the SAC is described in terms of only one 
spinor field, $\Psi\equiv \Psi_{S,1}$, which however is still coupled to the 
Majorana field $\gamma$.   After some algebra, we obtain the effective action
\begin{eqnarray}\label{Seff}
S_{\rm eff} &=& \int_0^\beta d \tau d \tau' \Biggl\{ 
\bar \Psi(\tau) K^{-1}(\tau-\tau') \Psi(\tau')  \\
&+& \Phi^T(\tau) \left[ \frac12 \delta(\tau-\tau')  \partial_{\tau'} -
\lambda_2^2 P_\uparrow g_2(\tau - \tau') P_\uparrow \right] 
\Phi(\tau')  \nonumber \\ \nonumber
&+&  \Biggl [ 
\bar \Psi(\tau) \Bigl( \lambda_1 e^{i\phi_1/2}\delta(\tau-\tau')  \\ 
&-& \lambda_2 
e^{i\phi_2/2} t_0  \tau_z e^{i\tau_z\phi/2} 
g_2(\tau-\tau') \Bigr) P_\uparrow \Phi(\tau') + {\rm h.c.} \Biggr] \Biggr\} ,
\nonumber
\end{eqnarray}
where  the operator $P_\uparrow = 
(\tau_0 + \tau_z)/2$ projects a Nambu spinor to its 
spin-$\uparrow$ component. 
Moreover, we have defined an effective GF in Nambu space with frequency components 
\begin{equation}
K^{-1}(\omega) = g^{-1}_1(\omega) - t_0^2 \tau_z e^{i\tau_z\phi/2} 
g_2(\omega) e^{-i\tau_z\phi/2} \tau_z,
\end{equation}
and the TS lead has been represented by the Majorana-Nambu spinor
\begin{equation}\label{Phi}
\Phi (\tau) = \frac{1}{\sqrt{2}} \left( \begin{array}{c} 1 \\ 1 \end{array}
 \right) \gamma (\tau) = \tau_x \Phi^\ast(\tau) .
\end{equation}
We note in passing that Eq.~(\ref{Seff}) could at this point  be generalized to 
include continuum states in the TS wire. To that end, one has to (i) replace $\Phi \to
 (\psi, \psi^\dagger)^T$, where $\psi$ is the boundary fermion of 
 the effectively spinless TS wire,
and (ii) replace $\delta(\tau-\tau')  \partial_{\tau'} \to G^{-1}(\tau-\tau')$
with $G$ in Eq.~(\ref{G}).
Including bulk TS quasiparticles becomes necessary for small values of the proximity gap,
$\Delta_p \ll \Delta$, and/or when studying nonequilibrium applications within 
a Keldysh version of our formalism.

In any case, after neglecting the above-gap TS continuum quasiparticles,  
the partition function follows with $S_{\rm eff}$ in Eq.~(\ref{Seff}) 
in the functional integral representation
\begin{equation}
Z = \int {\cal D} [\bar \Psi, \Psi, \gamma] e^{- S_{\rm eff}}\equiv e^{-\beta F(\phi_1,\phi_2)}.
\end{equation}
As before, the Josephson current through S lead no.~$j$ then follows from 
the free energy via $I_j = (2 e / \hbar) \partial_{\phi_j} F$.
The supercurrent flowing through the TS wire is then given by 
\begin{equation}\label{its}
I_{\rm TS} = -( I_1+I_2),
\end{equation}
as dictated  by current conservation.

\subsubsection{Atomic limit}

In order to get insight into the basic physics, we now analyze  in detail
the atomic limit, where $\Delta$ represents the largest energy scale
of interest and hence the dynamics is confined to the subgap region.
In this case, we can approximate
 $\sqrt{\Delta^2 + \omega^2} \approx \Delta$. After the rescaling 
 $\Psi \to \sqrt{\Delta/(1 + t_0^2)} \Psi$ in Eq.~(\ref{Seff}), 
we arrive at an effective action, $S_{\rm eff}\to S_{\rm at}$, valid in the atomic limit, 
\begin{eqnarray}\nonumber
S_{\rm at} &=& \int_0^\beta d \tau \Biggl\{
\frac12 \gamma \partial_\tau\gamma + \bar \Psi
\Bigl[ \partial_\tau +  \Delta \cos(\phi/2) \tau_x  + \\ \label{Sat}
&& \quad + r \Delta \sin(\phi/2) \tau_y  + B_z \tau_0 \Bigr] \Psi + \\ \nonumber 
&&\quad+\frac{1}{\sqrt{2}} \sum_{\sigma = \uparrow, \downarrow}
\left( \lambda_\sigma \psi_\sigma^\dagger - {\rm h.c.} \right) \gamma
\Biggr\},
\end{eqnarray}
where $r = \sqrt{1 - {\cal T}}$ is the reflection amplitude of the SAC, see Eq.~(\ref{tau0}).
We recall that $\Psi=(\psi_\uparrow,\psi_\downarrow^\dagger)^T$, see Eq.~(\ref{PsiSj}).
Moreover, we define the auxiliary parameters
\begin{eqnarray} \label{lambdas}
\lambda_\uparrow &=& \lambda_1 \sqrt{(1 + r) \Delta / 2} \, e^{i \phi_1/2} ,\\ \nonumber
\lambda_\downarrow &=& -\lambda_2 \sqrt{(1 - r) \Delta / 2} \, e^{-i \phi_2/2} , \\ \nonumber
B_z &= & \left( \frac{1+r}{2} b_1 + \frac{1-r}{2} b_2 \right) \Delta.
\end{eqnarray}
The parameters $b_{1,2}$ in Eq.~(\ref{ggg}) thus effectively generate the Zeeman scale $B_z$ in Eq.~(\ref{lambdas}).

As a consequence of the atomic limit approximation, the action $S_{\rm at}$ in Eq.~(\ref{Sat})
is equivalently expressed in terms of the effective Hamiltonian
 \begin{eqnarray}\label{Hat}
H_{\rm at} & = & \sum_{\sigma = \uparrow, \downarrow = \pm} 
\sigma B_z \psi_\sigma^\dagger \psi_\sigma +
\left( \delta_A \psi_\uparrow^\dagger \psi_\downarrow^\dagger + {\rm h.c.} \right)
\\ \nonumber &+&
\frac{1}{\sqrt{2}} \sum_\sigma
\left( \lambda_\sigma \psi_\sigma^\dagger - {\rm h.c.} \right) \gamma,
\end{eqnarray}
where we define
\begin{equation}
\delta_A(\phi) = \Delta \left[ \cos(\phi/2) -i r \sin(\phi/2) \right].
\end{equation}
For a SAC decoupled from the TS wire and taken at zero field ($B_z=0$), 
the ABS energy follows from Eq.~(\ref{Hat}) in the standard form \cite{Alvaro2011}
\begin{equation}\label{ea}
E_A(\phi) = |\delta_A| = \Delta \sqrt{1 - {\cal T}\sin^2(\phi/2)}.
\end{equation} 
We emphasize that $H_{\rm at}$ neglects TS continuum quasiparticles as well as 
all types of quasiparticle poisoning processes. 
Let us briefly pause in order to make two remarks. First,
we note that the Majorana field $\gamma=(f_\uparrow+f_\uparrow^\dagger)/\sqrt2$, 
see Eq.~(\ref{fdef}), couples to both spin modes $\psi_\sigma$ in Eq.~(\ref{Hat}).  
The coupling $\lambda_\downarrow$ between $\gamma$ and 
the spin-$\downarrow$ field in the SAC, $\psi_\downarrow$, is  
generated by crossed Andreev reflection processes, where a Cooper pair in lead S2 splits
according to
$\psi^\dagger_{2, \uparrow} \psi^\dagger_{2, \downarrow} \to f_\uparrow^\dagger 
 \psi^\dagger_{1, \downarrow}$, plus the conjugate process.
Second, we observe that $H_{\rm at}$ is 
invariant under a particle-hole transformation, amounting to the replacements 
$\psi_\sigma \to \psi_\sigma^\dagger$ and 
$f_\uparrow \to f_\uparrow^\dagger$, along with $B_z \to - B_z$ and $\phi_j \to 2 \pi - \phi_j$.

We next notice that with $n_\sigma=\psi_{\sigma}^\dagger
\psi_\sigma^{}=0,1$ and $n_f=f_\uparrow^\dagger f_\uparrow^{}=0,1$,
the total fermion parity of the junction,
\begin{equation}
{\cal P}_{\rm tot} = (-1)^{n_f+n_\uparrow+n_\downarrow} = \pm 1,
\end{equation}
is a conserved quantity, $\left[ {\cal P}_{\rm tot} , H_{\rm at} \right]_- = 0$.
Below we restrict our analysis to the even-parity sector ${\cal P}_{\rm tot} = +1$, 
but analogous results hold for the odd-parity case.
The corresponding Hilbert subspace is spanned by four states,
\begin{equation}\label{nnn}
| n_\uparrow, n_\downarrow, n_f \rangle = 
\left( \psi_\uparrow^\dagger \right)^{n_\uparrow}
\left( \psi_\downarrow^\dagger \right)^{n_\downarrow} 
\left( f_\uparrow^\dagger \right)^{n_f} | 0 \rangle ,
\end{equation}
where $(n_\uparrow, n_\downarrow, n_f) \in \left\{ (0,0,0), (1,1,0), (1,0,1), (0,1,1) \right\}$
and $|0\rangle$ is the vacuum state. In this basis, the Hamiltonian (\ref{Hat}) has the 
 matrix representation
\begin{equation}\label{h}
{\cal H}_{\rm at}(\phi_1, \phi_2) = 
 \left( \begin{array}{cccc} 0 & \delta_A^\ast & \lambda_\uparrow^\ast/2 & \lambda_\downarrow^\ast/2 \\
\delta_A & 0 & \lambda_\downarrow/2 & -\lambda_\uparrow/2 \\
\lambda_\uparrow/2 & \lambda_\downarrow^\ast/2 & B_z & 0 \\
\lambda_\downarrow/2 & - \lambda_\uparrow^\ast/2 & 0 & -B_z
\end{array} \right) .
\end{equation}
The even-parity ground state energy, $E_G^{\rm (e)} = {\rm min} (\varepsilon)$, follows as the
smallest root of the quartic equation 
\begin{equation}\label{quartic}
{\rm det} \left({\cal H}_{\rm at} - \varepsilon\right) = 0.
\end{equation}

In order to obtain simple results, let us now consider the special case $\lambda_2 = 0$, 
where the TS wire is directly coupled to lead S1 only, see Fig.~\ref{fig1}(b).
In that case, we also have $\lambda_\downarrow=0$, see Eq.~(\ref{lambdas}), 
and Eq.~(\ref{quartic}) implies the four eigenenergies $\pm \varepsilon_\pm$ with
\begin{eqnarray}\label{epspm}
\varepsilon_\pm & =& 
\frac{1}{\sqrt2} \Biggl( E_A^2 + B_z^2 +  \frac12 |\lambda_\uparrow|^2  \\ \nonumber
&\pm&
\sqrt{\left( E_A^2 - B_z^2 \right)^2 +  |\lambda_\uparrow|^2 \left( E_A^2 + B_z^2 \right)} \Biggr)^{1/2} ,
\end{eqnarray}
with $|\lambda_\uparrow|^2=\lambda^2_1 (1 + r) \Delta / 2$, see Eq.~(\ref{lambdas}).
The ground-state energy is thus given by $E_G^{\rm (e)} = - \varepsilon_+$. 
Since $E_G$ depends on the phases $\phi_{1,2}$ only via the Andreev level energy 
$E_A(\phi)$ in Eq.~(\ref{ea}), 
the Josephson current through the SAC is given by 
\begin{equation}\label{sac1}
I_1 = -I_2 = \frac{2e}{\hbar} \partial_\phi E_G^{\rm (e)}= - \frac{2e}{\hbar}\partial_\phi
\varepsilon_+ .
\end{equation}
Note that Eq.~(\ref{its}) then implies that no supercurrent flows into the TS wire.

Next we observe that in the absence of the TS probe ($\lambda_1 = 0)$, 
the even and odd fermion parity sectors of the SAC, ${\cal P}_{\rm SAC}=(-1)^{n_\uparrow+
n_\downarrow}=\pm 1$, are decoupled, see Eq.~(\ref{h}), and
Eq.~(\ref{epspm}) yields $E_G^{\rm (e)} = - \max (E_A, |B_z|)$. 
Importantly, the Josephson current is therefore fully blocked if the ground state is in the 
${\cal P}_{\rm SAC}=-1$ sector, i.e., for $|B_z|>E_A(\phi)$.
For $\lambda_1 \neq 0$, however, ${\cal P}_{\rm SAC}$ is not  conserved anymore. 
This implies that the MBS can act as parity switch between the two  Andreev sectors
with parity ${\cal P}_{\rm SAC}=\pm 1$.
Near the level crossing point at $E_A \approx |B_z|$, i.e.,
assuming
$\left| E_A^2 - B_z^2 \right| \ll |\lambda_\uparrow|^2 \ll E_A^2 + B_z^2,$
we obtain
\begin{equation}\label{parityswitch}
\varepsilon_\pm \simeq \frac{1}{\sqrt2} 
\left( E_A^2 + B_z^2 \pm  \lambda_1 \sqrt{2(1+r)\Delta (E_A^2 + B_z^2)} \right)^{1/2} ,
\end{equation}
which implies a nonvanishing supercurrent through the SAC even in the field-dominated regime, $|B_z| > E_A$. The MBS therefore acts as a parity switch and 
leaves a trace in the CPR by lifting the supercurrent blockade.

\subsubsection{Another interpretation}

Interestingly, for $\lambda_2=\phi_2=0$, the S-TS-S setup in Fig.~\ref{fig1}(b) could also be viewed as a toy model for an S-TS junction, 
where the TS part corresponds to a spinful model.  In that analogy,
the Nambu spinor $\Psi_{S,1}$ stands for the S lead while the spinful TS wire is represented by (i) the
Nambu spinor $\Psi_{S,2}$ which is responsible for the residual $s$-wave pairing correlations, and (ii) by the  MF $\gamma$ (or, more generally, by the Kitaev-chain spinless boundary fermion $\psi$)
which encodes $p$-wave pairing correlations.
Moreover, $t_0$ and $\lambda_1$ should now be understood as spin-conserving phenomenological tunnel couplings 
acting in the $s$-$s$ and $s$-$p$ wave channels, respectively.
The phase difference across this effective S-TS junction is $\phi=\phi_1$ and
the net S-TS tunnel coupling is given by $\lambda=\sqrt{t_0^2 + \lambda_1^2}$.
Putting $\lambda_1=0$ in the topologically trivial phase of the TS wire,  the Josephson current
carried by Andreev states in the $s$-$s$ channel is blocked when
the ground state is in the odd parity sector of the SAC.  For $\lambda_1\ne 0$, the MBS-mediated 
switching between odd and even parity sectors will now be activated and thereby  
 lift the supercurrent blockade.  

\subsubsection{Conventional midgap level}

A similar behavior as predicted above for the MBS-induced parity switch between ${\cal P}_{\rm SAC}=\pm 1$ sectors could also be
expected from a conventional fermionic subgap state  tunnel-coupled to the SAC.  Such a subgap state may be represented, e.g.,
 by a single-level quantum dot in the Coulomb blockade regime.
In particular, for a midgap (zero-energy) level with the fermion operator $d$, 
the Hamiltonian $H_{\rm at}$ in Eq.~(\ref{Hat}) has to be replaced with
 \begin{eqnarray}\label{tildeHat}
\tilde H_{\rm at}& =& \sum_{\sigma = \uparrow, \downarrow = \pm} \sigma B_z
 \psi_\sigma^\dagger \psi_\sigma +
\left( \delta_A \psi_\uparrow^\dagger \psi_\downarrow^\dagger + {\rm h.c.} \right) +
\\ \nonumber &+&
\sum_\sigma \left( \lambda_\sigma \psi_\sigma^\dagger d + {\rm h.c.} \right) .
\end{eqnarray}
In the even total parity basis (\ref{nnn}), the matrix representation of 
the Hamiltonian is then instead of Eq.~(\ref{h}) given by
\begin{equation}
\tilde {\cal H}_{\rm at}(\phi_1, \phi_2) =
\left( \begin{array}{cccc} 0 & \delta_A^\ast & 0 & 0 \\
\delta_A & 0 & \lambda_\downarrow & -\lambda_\uparrow \\
0 & \lambda_\downarrow^\ast & B_z & 0 \\
0 & - \lambda_\uparrow^\ast & 0 & -B_z
\end{array} \right) .
\end{equation}
Assuming $|\lambda_\uparrow| = |\lambda_\downarrow| \equiv \lambda$, 
Eq.~(\ref{quartic}) then yields the eigenenergies $\pm \varepsilon_\pm$ with
\begin{eqnarray}\label{eppp}
\varepsilon_\pm &=& \frac{1}{\sqrt2} \Biggl( E_A^2 + B_z^2 + 2 \lambda^2  \\
&& \nonumber \pm
\sqrt{\left( E_A^2 - B_z^2 \right)^2 + 4 \lambda^2 
\left( E_A^2 + B_z^2  + \lambda^2 \right) } \Biggr)^{1/2} .
\end{eqnarray}
Remarkably, the ABS spectra in Eqs.~(\ref{eppp}) and (\ref{epspm}) are rather similar for 
$\lambda^2 \ll \max (E_A^2, B_z^2)$.
However, the MBS will automatically be located at zero energy and thus represents
a generic situation.

\section{Conclusion}

We close this paper by summarizing our main findings.  We have studied the Josephson effect in different setups involving both conventional
$s$-wave BCS superconductors (S leads) and topologically nontrivial 1D $p$-wave superconductors (TS leads) with Majorana end states.
The TS wires have been described either by a spinless theory applicable in the deep topological regime, which has the advantage of allowing
for analytical progress but makes it difficult to establish contact to experimental control parameters, or by a spinful nanowire model as suggested
in Refs.~\cite{Lutchyn2010,Oreg2010}.  We have employed a unified imaginary-time Green's function approach to analyze the equilibrium properties 
of such devices, but a Keldysh generalization is straightforward and allows one to study also nonequilibrium applications. 

For S-TS tunnel junctions, we find that in the topological phase of the TS wire, the supercurrent is mainly carried by above-gap continuum contributions.   
We confirm the expected supercurrent blockade \cite{Zazunov2012} in the deep topological regime (where the spinless theory is fully valid and thus no residual $s$-wave pairing exists), while for realistic parameters, a small but finite critical current is found.  To good approximation, the Josephson current obeys the usual $2\pi$-periodic sinusoidal 
current-phase relation.  The dependence of the critical current on the bulk Zeeman field driving the TS wire through the topological phase transition shows a kink-like feature at the critical value,
which is caused by a sudden drop of the Andreev state contribution. 

The supercurrent blockade in the deep topological phase could  be lifted by adding a magnetic impurity to the junction, also allowing for the presence of a local magnetic field ${\bf B}$.
Such a magnetic impurity arises from a spin-degenerate quantum dot (QD), and we have studied the corresponding S-QD-TS problem for both the spinless and the spinful TS wire model.
Based on analytical results valid in the cotunneling regime as well as numerical results within the mean-field approximation, we predict $\varphi_0$-junction behavior (anomalous Josephson effect) for the current-phase relation when the TS wire is in the topological phase.   

As a final example for devices combining conventional and topological superconductors, we have shown that S-TS-S devices allow for a Majorana-induced parity switch between Andreev 
state sectors with different parity in a superconducting atomic contact.  This observation could be useful for future microwave spectroscopy experiments of Andreev qubits in such contacts.

\begin{acknowledgements}
We acknowledge funding by the Deutsche Forschungsgemeinschaft (Grant No.~EG 96/11-1)
and by the Spanish MINECO through Grant No.~FIS2014-55486-P
and through the ``Mar\'{\i}a de Maeztu'' Programme for Units of Excellence in R\&D (MDM-2014-0377). 
\end{acknowledgements}

%\bibliography{bib-beilstein}

\end{document}